\documentclass[preprint2]{aastex}                
\renewcommand\ion[2]{#1\,{\sc{\romannumeral #2}}} 
\newcommand{\copernicus}{{\it Copernicus}}

\newcommand{\fuse}{{\it FUSE}}

\newcommand{\ha}{H$\alpha$}

\newcommand\ilevel[2]{#1\,$^{#2}$}

\newcommand{\kms}{km\,s$^{-1}$}

\newcommand{\mdot}{$\dot{M}$}

\newcommand{\mdotq}{$\dot{M} q$}

\newcommand{\msunpyr}{${\rm M}_\sun / {\rm yr}$}
\newcommand{\rsun}{$R / R_\sun$}

\newcommand{\teff}{$T_{\rm eff}$}

\newcommand{\vinf}{${v}_\infty$}
\newcommand{\vsini}{$v \sin i$}
\newcommand\tna{\tablenotemark{a}}
\newcommand\tnb{\tablenotemark{b}}
\slugcomment{Accepted by ApJ October 4, 2005} 
\shortauthors{Fullerton et al.}                                                %
\shorttitle{Discordant Mass-Loss Estimates for O Stars}                        %
\begin{document}

\title{The Discordance of Mass-Loss Estimates \\for Galactic O-Type Stars}

\author{A.~W. Fullerton\altaffilmark{2}}
\affil{Dept. of Physics \& Astronomy,
       University of Victoria,
       P.O. Box 3055,
       Victoria, BC, {V8W 3P6}, Canada.}
\altaffiltext{2}{Postal Address: 
                 Dept. of Physics \& Astronomy,
                 The Johns Hopkins University,
		 3400 N. Charles Street,
		 Baltimore, MD 21218.}
\email{awf@pha.jhu.edu}

\author{D.~L. Massa}
\affil{SGT, Inc,
       NASA's Goddard Space Flight Center,
       Code 681.0,
       Greenbelt, MD 20771.}
\email{massa@taotaomona.gsfc.nasa.gov}

\and

\author{R.~K. Prinja}
\affil{Dept. of Physics \& Astronomy,
       University College London,
       Gower Street, 
       London {WC1E 6BT},
       UK.}
\email{rkp@star.ucl.ac.uk}
\begin{abstract}
We have determined accurate values of the product of the mass-loss rate
and the ion fraction of {\ilevel{P}{4+}}, {\mdotq(\ilevel{P}{4+})},
for a sample of 40 Galactic O-type stars by fitting stellar-wind profiles
to observations of the {\ion{P}{5}} resonance doublet obtained with
{\fuse}, {\it ORFEUS/BEFS}, and {\copernicus}.
When {\ilevel{P}{4+}} is the dominant ion in the wind (i.e.,
{0.5 $\la$ $q$(\ilevel{P}{4+}) $\le$ 1}), {\mdotq(\ilevel{P}{4+})} approximates
the mass-loss rate to within a factor of $\la$2.
Theory predicts that \ilevel{P}{4+} is the dominant ion in the winds of 
O7--O9.7 stars, though an empirical estimator suggests that the range from O4--O7 may 
be more appropriate.
However, we find that the mass-loss rates obtained from {\ion{P}{5}} wind 
profiles are systematically smaller than those obtained from fits to {\ha} 
emission profiles or radio free-free emission by median factors of 
$\sim$130 (if {\ilevel{P}{4+}} is dominant between O7 and O9.7) or 
$\sim$20  (if {\ilevel{P}{4+}} is dominant between O4 and O7).
These discordant measurements can be reconciled if the winds of O
stars in the relevant temperature range are strongly clumped on small 
spatial scales.
We use a simplified two-component model to investigate
the volume filling factors of the denser regions. 
This clumping implies that mass-loss rates determined from
``$\rho^2$" diagnostics have been systematically over-estimated by
factors of 10 or more, at least for a subset of O stars.
Reductions in the mass-loss rates of this size have important
implications for the evolution of massive stars and quantitative estimates 
of the feedback that hot-star winds provide to their interstellar 
environments.
\end{abstract}

\section{Introduction}\label{intro}

The stellar winds from OB-type stars exert strong, dynamic influences on 
the evolution of massive stars and their interstellar environments.  
From the perspective of stellar evolution, continuous mass-loss via these 
outflows alters both the path of the star through the H-R diagram and the 
rate at which it is traversed.  
The continual shedding of the outer layers of the atmosphere also causes 
systematic changes in the chemical composition of the photosphere, particularly
in the presence of strong, rotationally-induced mixing.  
Additionally, the injection of chemically enriched material, momentum, and 
energy from the winds of hot, massive stars into their surroundings is an 
integral part of the ``stellar feedback" cycle, which mixes the 
local interstellar medium and ultimately drives the chemical evolution of a 
galaxy.

In view of their fundamental astrophysical importance, it is clear that the
properties of these outflows must be well determined.  
Nowadays, it is understood that hot-star winds are driven by the transfer of 
momentum from the stellar radiation field through scattering in metallic 
resonance lines.  
However, {\it ab initio}\/ models still require numerous assumptions in order
to provide testable predictions of wind parameters.  
Consequently, empirical confirmation of the theory remains essential.  
For this purpose, the three most useful diagnostics of mass loss are, in order 
of increasing sensitivity: 
 (a)~free-free continuum emission at radio wavelengths;
 (b)~{\ha} line emission; and 
 (c)~ultraviolet (UV) resonance-line absorption.
These diagnostics sample different parts of the wind, from the dense, 
near-star, rapidly accelerating region (\ha) to the very distant, 
rarefied, constant velocity regions (free-free radio emission) and 
essentially everywhere between (UV resonance lines).   
Their physical mechanisms have different dependencies on the local density 
(emission $\propto \rho^2$; absorption $\propto \rho$), while the measurable 
properties associated with each suffer different degrees of contamination 
from photospheric or other processes, and require different ancillary 
knowledge of, e.g., the excitation or ionization conditions in the wind, 
the velocity structure of the wind, or the distance to the star.

The need for additional information concerning the ionization 
structure of the wind has traditionally been a stumbling block for 
determinations of the mass-loss rates ({\mdot}) from the wind profiles 
of UV resonance lines.  
The problem arises because the strength of these profiles is determined 
by the radial optical depth of the wind, 
 $\tau_{rad} \propto \dot{M} q_i A_E$, 
where $A_E$ and $q_i$ are the abundance of element $E$ and its ionization 
fraction for stage $i$.  
Thus, wind-profile modeling of a dominant ion ($q_i\sim1$) of known 
abundance is required to estimate $\dot{M}$\/ directly.
Unfortunately, the lines of dominant ions are usually saturated, 
particularly those for cosmically abundant elements (e.g., C, N, O) in 
winds that are sufficiently dense to provide reliable determinations of
{\mdot} from ``density squared" [hereafter ``$\rho^2$" and \mdot($\rho^2$)] 
diagnostics.
As a result, only UV wind lines from trace ionization species with 
$q_i \lesssim 10^{-3}$ can typically be used to measure {\mdotq} for massive 
winds, and values of {\mdot} estimated from these measurements are compromised 
by the lack of {\it a priori} information needed to estimate $q_i$ accurately;
see \citet{Lamers99} for additional discussion of this problem.
Thus, even though UV wind lines are the most sensitive indicators of 
{\mdot} in early-type stars, empirical mass-loss rates have relied primarily 
on measurements {\mdot($\rho^2$)} from free-free radio emission
[{\mdot(radio)}] and {\ha} line emission [{\mdot(\ha)]; 
see, e.g., \citet{Lamers93} or \citet{Puls96}.

Although mass-loss rates determined from radio measurements are usually
regarded as the most reliable, comparatively few OB-type stars have 
been detected owing to the inherent weakness of free-free emission and 
its ``$\rho^2$" dependence, which requires stars to be relatively nearby.  
In contrast, {\ha} emission, which is also a ``$\rho^2$" processes, 
can be easily observed in distant objects.
However, because its strength depends on the details of the {wind--photosphere}
interface, it is more difficult to model, although the current consensus is 
that mass-loss rates determined from {\ha} are quite accurate 
{\citep{Puls96}}.

Access to the far-ultraviolet region of the spectrum provided by the
{\it Far Ultraviolet Spectroscopic Explorer} (\fuse) affords a renewed
opportunity to examine the use of resonance-line diagnostics 
for the determination of {\mdot}, and to perform consistency checks
on measurements of {\mdot($\rho^2$)}.
For this purpose, the {\ion{P}{5}}~$\lambda\lambda$1118, 1128 doublet plays
a pivotal role, primarily because {\ilevel{P}{4+}} is expected to be the 
dominant ion the winds of {\em some} O stars.
This expectation is based on naive energetic considerations, 
morphological trends observed in the spectra of O stars, and 
detailed modeling.
From a diagnostic perspective, an equally important attribute of 
{\ilevel{P}{4+}} is that, despite its high ionic abundance, 
the {\ion{P}{5} $\lambda\lambda$1118, 1128} doublet is rarely saturated 
because the cosmic abundance of P is so small.  
On the other hand, this also means that the wind contribution to the doublet
is only measurable in stars with massive winds.  
Finally, in contrast to C, N, and O, P is not produced by H-burning; 
consequently its atmospheric abundance does not change substantially over a 
stellar lifetime or as a function of the vigor of rotational mixing.  
Thus, star-to-star changes in its abundance should not contribute to the 
cosmic scatter in {\mdot} determinations.

These properties make the {\ion{P}{5}} resonance doublet a powerful
mass-loss diagnostic for a range of O-star temperature classes, though
we expect the detection of wind profiles in the {\ion{P}{5}} resonance
doublet to be biased toward stars with more massive winds. 
If we simply assume that $q$(\ilevel{P}{4+}) approaches unity in stellar
winds {\em somewhere} in the O star range, then values of 
\mdotq(\ilevel{P}{4+}) should agree with both {\mdot(radio)} and 
{\mdot(\ha)} for at least {\em some} O stars.  
If concordance between these individually reliable mass-loss indicators is 
not found, then their different formation mechanisms will help to constrain
the reasons for the discrepancy.
\citet{Massa05} reported on a preliminary comparison between these different 
measures of {\mdot}, and the current paper expands and refines that work. 

In the present study, we have measured $\dot{M}q({\rm P}^{4+})$ for a sample 
of 40 well-studied Galactic O-type stars that have reliable radio or {\ha} 
mass-loss measurements available from the literature.  
This sample is described  in \S\ref{sample}.  
New {\fuse} observations and wind-profile analysis of all {\ion{P}{5}} 
resonance doublets are described in \S\ref{obs} and \ref{uv_mdot}, 
respectively.  
The values of {\mdotq} derived for {\ilevel{P}{4+}} are compared with the 
values of {\mdot($\rho^2$)} in \S\ref{results}, and large discrepancies are 
found.  
Possible reasons for this discordance are discussed in \S\ref{discuss},
before the main conclusions from this study are summarized in \S\ref{conclude}.

\section{The Sample}\label{sample}

Table~\ref{targets} lists the sample of Galactic O-type stars used in this 
investigation.
The sample is defined by the availability of
(a) reliably determined values of the fundamental stellar parameters;
(b) reliable determinations of {\mdot} from thermal radio emission 
    or {\ha} emission profiles; and
(c) far-UV spectra that include the {\ion{P}{5}} resonance lines.
For practical purposes, the first two criteria limit the selection 
of objects to those that have been well-studied over the past decade
by the methods of ``quantitative spectroscopy."  
These objects serve as a fundamental data set, and are the cornerstone for 
the calibration of the Wind-Momentum-Luminosity relation for Galactic O stars 
\citep{Puls96,Repolust04}.  
The sample spans all subclasses of O-type spectra, but is biased toward 
higher luminosity classes.  
The luminosity bias occurs because both  ``$\rho^2$" diagnostics and the 
{\ion{P}{5}} wind features are preferentially detected in the densest 
outflows, which are associated with more luminous objects.

\subsection{Stellar Parameters}

As described in \S\ref{uv_mdot}, determinations of 
{\mdotq(\ilevel{P}{4+})} from the {\ion{P}{5}} resonance lines requires 
knowledge of the stellar radius, {$R_\star$}; the projected rotational 
velocity, {\vsini}, which determines the width of the photospheric lines; 
and the terminal velocity of the stellar wind, {\vinf}.  
These parameters  and their origin are indicated in Table~\ref{targets}, 
along with the spectral type and the adopted values of {\teff}.  

To minimize possible systematic biases in the derived stellar parameters, 
the sample is subdivided into ``primary" (28 objects) and ``secondary" 
(12 objects) subsets according to how the parameters were determined.  
Most of the determinations for the primary data set come from the work of
\citet{Repolust04}, which provides the largest sample of homogeneously 
determined stellar parameters currently available for Galactic O stars.  
These are derived from fits made with the unified (i.e., 
photosphere $+$ wind), non-LTE, line-blanketed model atmosphere program 
{\tt FASTWIND} \citep{Puls05}.
The primary sample also contains objects analyzed by \citet{Markova04} 
with a more simplified approach, which relies in part on calibrations of 
fundamental parameters derived from the work of \citet{Repolust04}.  
Thus, the values of the derived stellar parameters for the primary sample are 
internally consistent and incorporate recent revisions to the temperature scale 
associated with O-type stars 
{\citep{Martins02,Crowther02,Bianchi02,Herrero02,Martins05}}.

Since similar fits with unified model atmospheres are not yet available for
stars in the secondary sample, the values of {\teff} and {\rsun} listed for
them in Table~\ref{targets} are based on the revised spectral-type calibrations
presented by \citet{Martins05}.
The parameters given for the ``observational scale" 
{\citep[i.e., Tables 4--6 of][]{Martins05}} were used to minimize possible
systematic differences with the primary sample.
Values for luminosity class II were estimated by linear interpolation
between luminosity classes I and III; and the parameters associated
with O9.5 supergiants were adopted for the three O9.7 supergiants in 
the secondary sample.
Of course, it would be very useful to determine the stellar parameters
of all the stars listed in Table~\ref{targets} through rigorously
uniform analysis. 

Table~\ref{targets} also notes which objects in the sample are known
or suspected binary stars, as designated in the Galactic O-Star Catalog
\citep{GOSCat04}.

\section{Far-Ultraviolet Spectra}

The far-UV spectra used to measure the {\ion{P}{5}} wind profiles of the 
targets in the sample were acquired by several observatories.
These are indicated in Table~\ref{obslog}, along with other salient details 
concerning the observations.  
Most of the spectra were obtained with {\fuse}, either as part of our Guest 
Investigator program E082, which was dedicated to measurements of {\ion{P}{5}},
or as a by-product of programs devoted to other topics.  
The processing of these data is described in \S\ref{obs}.

Owing to count-rate limits associated with the {\fuse} detectors, bright 
objects are difficult or impossible to observe.  
Consequently, {\fuse} cannot observe nearby, unreddened O stars.  
This is unfortunate, since these are generally the objects that have 
received the most attention from ground-based observers and for which radio 
mass-loss rates are available.  
For these bright stars, we have used archival far-UV spectra that were
obtained by {\copernicus} or the {\it Berkeley Extreme and Far-UV 
Spectrometer} ({\it BEFS}).  
These spectra were retrieved from the Multi-Mission Archive at the Space 
Telescope Science Institute (MAST).\footnote{The archiving of non-HST data at 
MAST is supported by the NASA Office of Space Science via grant NAG5-7584 and 
by other grants and contracts.  STScI is operated by the Association of 
Universities for Research in Astronomy, Inc., under NASA contract NAS5-26555.}
The data sets for these instruments are indicated in Table~\ref{obslog}, and 
were used without further manipulation.

\subsection{ {\fuse} Observations and Reduction}\label{obs}

The {\fuse}\/ observatory consists of four aligned, prime-focus telescopes
and Rowland-circle spectrographs that feed two photon-counting detectors
\citep{Moos00,Sahnow00}.
These four channels provide redundant coverage of the range between
$\sim905 - 1187$~{\AA}.
As indicated in Table~\ref{obslog}, most objects were observed through the 
large 30\arcsec $\times$ 30\arcsec (LWRS) aperture.  
The lone exception is HD~188209, which was observed through the narrow 
1.25\arcsec $\times$ 20\arcsec (HIRS) aperture as part of a test of 
observing techniques for bright objects.

The {\fuse} spectra were uniformly extracted and calibrated with 
CalFUSE version 2.4.2.
Subsequent processing used the shifts determined by cross-correlating
the positions of sharp interstellar features to align the spectra
extracted from individual exposures; resampled the spectra onto a uniform 
wavelength grid with steps of 0.1~{\AA}; and normalized the profiles to the 
local stellar continuum.  
Since normalization of LiF1B spectra acquired through the LWRS aperture
is complicated by an optical artifact known as ``the worm" 
\citep{Sahnow02}, we used the redundant and cleaner coverage of the 
{\ion{P}{5}} doublet provided by LiF2A spectra whenever possible.

\section{Mass-Loss Determinations}\label{mdot}

Determinations of the mass-loss rates for O stars traditionally assume that 
the wind is a spherically symmetric, steady, homogeneous, and monotonically 
expanding outflow.  
All the measurements discussed in this section depend on these underlying 
assumptions of this ``standard wind model.''

\subsection{{\ha} Measurements}\label{ha_mdot}

Most of the stars in the sample have {\mdot(\ha)} measurements.
The bulk of these were obtained from comparisons with {\tt FASTWIND} 
calculations of the entire optical spectrum by \citet{Repolust04}, which 
were also used to derive other fundamental stellar parameters.
Since these computations help to define the revised temperature scale, 
the values of {\mdot(\ha)} are completely consistent with it.

An important subset of objects have {\mdot(\ha)} determined from fits 
to only the {\ha} line profile by \citet{Markova04}.  
They used the quick but accurate method developed by \citet{Puls96}.  
These measurements rely in part on calibrations of stellar parameters as a 
function of spectral type that were based on the complete modeling of 
\citet{Repolust04}.
\citet{Markova04} show that their results are internally consistent
with the values derived from the more complete analysis.

Less reliable estimates of {\mdot(\ha)} were used for four stars
in the secondary sample: HD~45160, HD~46223, HD~164794, and HD~188001.
For these stars, {\mdot} was determined by
\citet[][their Table~11]{Puls96} from a re-analysis of the equivalent 
width measurements compiled by \citet{Lamers93}. 
These values are not of comparable quality to the other determinations used 
here, since they
(a) rely on equivalent widths rather than fits to the entire line profile; 
(b) are in some cases obtained from measurements of photographic plates, 
    which are inherently less precise than the CCD data used by 
    \citet{Repolust04} and \citet{Markova04}; and 
(c) were interpreted on the basis of a temperature scale that is 
    systematically hotter than the one used for the bulk of the stars 
    analyzed here.
Since three of these targets (HD~46150, HD~46223, HD~164794) only have upper 
limits for \mdotq(\ilevel{P}{4+}), the greater uncertainty inherent in their 
values of {\mdot(\ha)} does not affect the comparison significantly.
The fourth star, HD~188001, is the only program star in this group with 
detections of both \mdot(\ha) and \mdotq(\ilevel{P}{4+}).
Although the {\teff} of this star has been adjusted in Table~\ref{targets},
its value of {\mdot(\ha)} has not been rescaled.

\subsection{Measurements of Radio Free-Free Emission}\label{ff_mdot}

Mass-loss rates determined from radio free-free emission are available
for 40\% of the stars in the sample.  
Most of these have been taken from the compilation of \citet{Lamers93}, 
who provide a thorough discussion of this technique. 
These results are supplemented by the more recent detections of HD~14947 
and HD~190429A by \citet{Scuderi98}.  
Following the procedure adopted by \citet{Lamers93}, the 
uncertainties quoted by \citet{Scuderi98} have been increased by 
$\pm$0.10~dex to account approximately for uncertainties in the 
distance to the objects.

However, all these determinations pre-date the revisions to the 
effective temperature scale for O stars.
At first sight, {\mdot(radio)} seems immune from these changes, 
since {\teff} enters the expression for {\mdot} explicitly only very 
weakly as a logarithmic contribution to the Gaunt factor; see, e.g., 
equation~(5) of \citet{Lamers93}.
In fact, the resultant changes in $M_V$ and stellar luminosity generally
result in different estimates for the distance, $d$, to the various stars;
see, e.g., \citet{Martins05}.
Since {\mdot(radio)} $\propto d^{1.5}$, a systematic difference results
between {\mdot(\ha)} (which, except for HD~188001, are on the cooler 
temperature scale) and {\mdot(radio)}.

To account for this dependence, we have systematically adjusted the 
values of {\mdot(radio)}.
For stars in the primary sample, we used the value of $M_V$ quoted
by \citet{Repolust04} and \citet{Markova04}, together with 
photometry from the Galactic O-Star Catalog \citep{GOSCat04},
to estimate the distance modulus via the standard formula
\begin{equation}
DM = V - M_V - 3.1\,E(B-V)
\end{equation}
This leads to a revised estimate of the distance, $d_{new}$, and
a correction factor for the published value of {\mdot(radio)} of
$f_{corr}=(d_{new}/d_{old})^{1.5}$,
where $d_{old}$ was listed by \citet{Lamers93} or \citet{Scuderi98}.
The same approach was followed for stars in the secondary sample, though
in this case internally consistent values of $M_V$ were adopted from
the spectral-type calibration of \citet{Martins05}.
Table~\ref{revmdot} lists the correction factors that were applied
to make the measurements of {\mdot(radio)} consistent with the
measurements of {\mdot(\ha)}.

\subsection{{\ion{P}{5}} Wind-Profile Fitting}\label{uv_mdot}

The strength of the absorption trough of a P~Cygni wind profile at any 
normalized velocity $w = v/v_\infty$ (where {\vinf} is the terminal 
velocity of the wind) depends on the radial optical 
depth of material along the line-of-sight at that velocity, 
$\tau_{rad}(w)$, which is directly related to the column density.  

For the {\ion{P}{5}} resonance lines (whose basic properties are listed in 
Table~\ref{pfive}), $\tau_{rad}(w)$ for each program star was 
determined by using the ``Sobolev with Exact Integration" 
\citep[SEI;][]{Lamers87} wind-profile fitting technique with the 
modifications described by \citet{Massa03}.  
These modifications permit the relationship between optical depth and position
in the absorption trough to be an arbitrary function, and also allow 
approximately for the effects of interstellar absorption, whenever required.  
The analysis proceeded in three steps.

First, the parameters of the velocity law were defined.  
The velocity law, $w(x)$, was parameterized in terms of the usual 
$\beta$-law by
\begin{equation}\label{eq_vlaw}
   w(x)   = w_0 + ( 1 - 1/x )^\beta~~
\end{equation}  
where $w = v / v_\infty$ is the normalized velocity;
$x = r/R_\star$ is the radial distance in units of the stellar radius; 
and where the radial gradient is controlled by the parameter $\beta$, 
which is typically $\sim$1.  
The adopted values of {\vinf} are listed in Table~\ref{targets}.  
These were determined from the shape of strongly saturated 
P~Cygni profiles, typically the {\ion{C}{4}~$\lambda\lambda$1548, 1550} 
doublet, and included an estimate of any additional velocity dispersion 
(often referred to as ``turbulence") in the wind, which is usually required 
to reproduce the slope of the blue edge of the absorption trough.  
We found that adopting a value of {0.05\,\vinf} for this additional velocity 
dispersion was sufficient to model the {\ion{P}{5}} profiles.  
Similarly, we adopted $\beta = 1$ for all initial fits to the {\ion{P}{5}} 
doublet, and altered it by modest amounts only if significantly better fits 
resulted.  
The final values of $\beta$ are listed in Table~\ref{mdots}.

Second, the {\ion{P}{5}} doublet in each spectrum was fitted with the adopted 
velocity law by adjusting the optical depth in the radial direction, 
$\tau_{rad}(w)$, in $\sim$20 velocity bins distributed through the 
absorption trough.  
As explained in detail by \citet{Massa03}, determining the optical depth in 
absorption in a given bin also determines the contribution to blue- and 
red-shifted emission, so that a satisfactory fit to the entire profile is 
constructed by stepping through the absorption trough from its blue edge 
to line center.  
Underlying photospheric profiles were approximated by Gaussians, but since 
their influence is confined to low velocities (within $\pm$\vsini; see 
Table~\ref{targets} for the adopted values), their inclusion has little 
effect on the determination of $\tau_{rad}(w)$ for $w \gtrsim 0.2$,
{\em except} in cases where the wind profile is extremely weak.
A bigger problem is the blend of {\ion{P}{5}~$\lambda$1128} with a strong, 
excited transition of {\ion{Si}{4}} in the spectra of later O-type stars.  
In these cases, greater weight was given to the {\ion{P}{5}~$\lambda$1118} 
profile.  
Figures~\ref{PV_fits} and \ref{PV_O7fits} illustrate the quality 
of the fits for a selection of targets.
       
Third, these determinations of $\tau_{\rm rad}(w)$ were used to compute
$\dot{M} q({\rm P}^{4+}; w)$ for each star from the relation:
\begin{equation}\label{eq_mdotq}
  \dot{M} q ({\rm P}^{4+}; w) = 
  \left( \frac{m_e c}{\pi e^2}\right) \frac{4 \pi \mu m_H}
  {f_{ij} \lambda_0 A_P} R_\star v_{\infty}^2 x^2 w \frac{dw}{dx} \tau_{rad}(w) 
\end{equation}
where
$f_{ij}$ is the oscillator strength of the appropriate component of the
{\ion{P}{5}} resonance line; 
$A_P$ is the abundance of P relative to hydrogen by number 
(Table~\ref{pfive}); 
$\mu$ is the mean molecular weight of the plasma; and 
all other symbols have their usual meaning.  
In equation~(\ref{eq_mdotq}), we adopted $\mu= 1.34$ (appropriate for a 
completely ionized plasma of solar abundance), together with the values of 
$R_\star$ and {\vinf} listed in Table~\ref{targets}.  
The measurements of $\dot{M} q_i(w)$ were averaged over normalized velocities 
between 0.2 and 0.8 to obtain the values of \mdotq(\ilevel{P}{4+}) listed in 
Table~\ref{mdots}.

The errors associated with the derived $\tau_{rad}(w)$ depend on its 
magnitude.  
For $\tau_{rad}(w) \lesssim 0.05$, the error can be as large as a factor of 
two.  
Similarly, for $\tau_{rad}(w) \gtrsim 2.0$ the error can also be large, 
while for $\tau_{rad}(w) \sim 1.0$, it is on the order of 10\%. 
Consequently, we provide only upper limits for lines whose blue (stronger) 
component has an optical depth less than 0.05.  
None of the program stars has an optical depth in the red (weaker) component 
of the {\ion{P}{5}} doublet greater than 2.0, so there is no need to
define lower limits on $\tau_{rad}(w)$.
Typical errors in parameters derived from the integrated optical depths, 
such as \mdotq(\ilevel{P}{4+}), 
are $\lesssim 25$\%, and only quantities derived from the weakest lines 
have significantly larger errors.  
The value of $\beta$\/ can also affect $\tau_{rad}(w)$.  
However, $\beta$\/ is typically determined to within $\pm 0.2$, and 
uncertainties of this amount have very little effect on $\tau_{rad}(w)$ 
over the range $0.2 \leq w \leq 0.8$.

\section{Results}\label{results}

\subsection{Comparison of {\mdot($\rho^2$)} with {\mdotq(\ilevel{P}{4+})}}

The published values of {\mdot(radio)} [modified as described in
\S\ref{ff_mdot}] and {\mdot(\ha)} are listed in Table~\ref{mdots}.
They are compared with the average values of {\mdotq(\ilevel{P}{4+})} in 
Figure~\ref{cf_fig} as a function of spectral type, measurement technique, 
and status as members of the primary or secondary sample.  
When more than one {\mdot($\rho^2$)} measurement was available, 
{\mdot(radio)} was preferred, except in cases where the uncertainties 
associated the {\mdot(\ha)} measurement were substantially smaller.  
Since only upper limits on both {\mdot($\rho^2$)} and {\mdotq(\ilevel{P}{4+})} 
are available for HD~46150 and HD~217086, the values for these stars are not 
plotted in Fig.~\ref{cf_fig}.
The position of HD~149757 is not indicated either, since its very
small value of {\mdotq(\ilevel{P}{4+})} lies beyond the limits of the plot.

Several things are apparent from Figure~\ref{cf_fig}.  
First, {\mdotq(\ilevel{P}{4+})} is systematically smaller than  
{\mdot($\rho^2$)} by substantial amounts. 
Second, the size of the deviation does not depend on whether the fiducial 
{\mdot($\rho^2$)} was {\mdot(radio)} or {\mdot(\ha)}.
Third, any biases in the determinations of {\mdot} for stars in the 
secondary sample are much smaller than the systematic deviations from 
the line indicating a 1--1 correlation.
Finally, a strong dependence on binarity is not evident.  

However, systematic trends in the distribution of deviations with spectral 
class are apparent.  
The sample can be divided into three parts:
\begin{enumerate}
  
  \item The mid O-type stars (O4 -- O7.5) with strong wind features 
        (i.e., those with {$\log$ \mdotq(\ilevel{P}{4+}}) $\ge -8$) exhibit
	the smallest deviations from the values of {\mdot(\ha)} or
	{\mdot(radio)}.
  
  \item The earliest (O2--O3.5) and latest (O8--O9.7) O stars exhibit 
        systematically larger deviations from the 1--1 correlation line.
  
  \item The largest deviations belong to a group of five mid-O dwarfs and 
        giants that have {\mdot(\ha)} measurements, and only upper 
	limits for {\mdotq(\ilevel{P}{4+})}.  
	These stars are (from left to right in Fig.~\ref{cf_fig}:
        HD~42088 (O6.5 V), HD~46223 [O4 V((f+))], HD~47839 [O7 V((f))]
	HD~217086 (O7 Vn), and HD~203064 [O7.5 III:n((f))].

\end{enumerate}	

Figure~\ref{pv_teff} presents an alternate representation of this 
discrepancy in terms of an empirical estimate of the mean value of 
{$q$(\ilevel{P}{4+})}, 
\begin{equation}
q_{\rm est} \equiv \dot M\,q({\rm P^{4+}}) / \dot M( \rho^2 )~.
\end{equation}
In Fig.~\ref{pv_teff}, $q_{\rm est}$ is plotted as a function of {\teff}
for different luminosity classes.
Although expected to be near unity over some range of {\teff}, it
is never more than $\sim$0.11 for supergiants, and becomes progressively 
smaller for less luminous stars (0.06 for giants; 0.04 for dwarfs).
Thus, {\mdot($\rho^2$)} is systematically larger than 
{\mdotq(\ilevel{P}{4+}), as also indicated in Fig.~\ref{cf_fig}.
Since the initial expectation was that {\mdotq(\ilevel{P}{4+})}
should be nearly equal to {\mdot($\rho^2$)} for at least some O-type stars,
the size and systematic nature of the deviations requires explanation.

\subsection{For Which Stars is {\ilevel{P}{4+}} Dominant?}\label{p5range}

A crucial assumption implicit in our emphasis on {\ion{P}{5}} is that
{$q$(\ilevel{P}{4+}) $\approx$ 1} for O stars spanning some range
of {\teff}.
Apart from naive energetic considerations, this assumption is supported
by detailed model atmosphere calculations.
In particular, smooth-wind models computed with {\tt FASTWIND} indicate
that this temperature range is $\sim$31 -- 34~kK\footnote{We are grateful 
to R.-P. Kudritzki and M. A. Urbaneja for providing us with a grid of 
unpublished {\tt FASTWIND} models.}, which corresponds to spectral
types between O9.7 -- O7.5 for supergiants and O9.5 -- O8.5 for dwarfs.

However, the distribution of the empirical estimates of {$q$(\ilevel{P}{4+})
in Fig.~\ref{pv_teff} suggests that the appropriate range of
{\teff} might be shifted to hotter temperatures.
For more luminous objects, $q_{\rm est}$ exhibits a broad maximum 
between {\teff} $\sim$34 -- 40~kK (i.e., O7.5 -- O4; mid-range O stars).
A peak near $\sim$41~kK may also be indicated for dwarfs, though the 
behavior of {$q_{\rm est}$} is not very clear for these objects.

A simple interpretation of the distribution of {$q_{\rm est}$} with
{\teff} is that it traces the rise and fall of \ilevel{P}{4+} as the 
dominant ion in the wind.  
In this case, {{\ilevel{P}{3+}} would be the dominant ion for
{\teff $\lesssim$ 34~kK}, while the balance shifts in favor
of {{\ilevel{P}{5+}} for {\teff $\gtrsim$ 40~kK}.
For both extremes, $q$({\ilevel{P}{4+}) $\ll$ 1, as implied by 
Fig.~\ref{pv_teff}.
However, this straightforward interpretation is complicated by the
reliance of {$q_{\rm est}$} on {\mdot($\rho^2$)}.
For example, biases in the determination of {\mdot($\rho^2$)}
will also bias {$q_{\rm est}$}.
Evidently a more systematic modeling effort will be required to determine
the temperature range for which {\ilevel{P}{4+}} is the dominant ion.

Irrespective of this uncertainty in the appropriate range of {\teff}, 
we conclude that there is a large, systematic discrepancy between 
{\mdotq(\ilevel{P}{4+})} and the ``$\rho^2$" determinations, {\mdot(\ha)} 
and {\mdot(radio)}, for some O stars.

\begin{enumerate}

\item If, as predicted by standard {\tt FASTWIND} models, {\ilevel{P}{4+}} is 
      dominant in the winds of O7.5 -- O9.7 stars, then Table~\ref{mdots} 
      indicates that the median discrepancy for the subset of 15 luminous 
      (i.e., non-dwarf) stars with solid detections corresponds to 
      {\mdot($\rho^2$)/{\mdotq(\ilevel{P}{4+})}} = 129, with minimum and 
      maximum values of 17 and 501 for HD~24912 and HD~209975, respectively.
      
\item If, as suggested by a straightforward interpretation of 
      the behavior of {$q_{\rm est}$} in Fig.~\ref{pv_teff}, 
      {\ilevel{P}{4+}} is dominant for stars with spectral classes 
      between O4--O7.5, the median discrepancy for the subset of 13 luminous 
      stars with solid detections corresponds to
      {\mdot($\rho^2$)/{\mdotq(\ilevel{P}{4+})}} = 20, with
      minimum and maximum values of 9 and 245 for 
      HD~66811 and HD~15558, respectively.
      
\end{enumerate}

\subsection{Stars with weak winds}

The extremely large discrepancies for the mid O-type stars with weak winds 
(i.e., the class III -- V stars) are an interesting special case, which 
might result from systematic measurement errors in {\mdot(\ha)}. 
None of these targets have obvious wind profiles in {\ion{P}{5}}, so only 
upper limits for {\mdotq(\ilevel{P}{4+})} can be estimated.  
Similarly, none of these stars are detected in the radio, and none exhibit 
{\it bona fide} {\ha} emission profiles.  
As a result, these determinations of {\mdot(\ha)} depend sensitively on 
estimates of the degree to which the underlying photospheric profiles
are partially filled by wind emission, which in turn relies 
critically on the accuracy of the photospheric models.
Such wind contamination is clearly evident in the {\ha} profiles of 
HD~203064 {\citep{Repolust04}}, though emission from a circumstellar disk 
is not precluded for this rapid rotator.  
HD~217086 is also a rapid rotator, though in this case partial filling by 
emission is not evident, and only an upper limit on {\mdot} was determined 
{\citep{Repolust04}}.  
Emission is less evident in the {\ha} profiles of HD~47839, which might 
instead be weakly contaminated by the spectrum of its late-O companion 
{\citep{Gies93}}.  

In any case, if the determinations of {\mdot(\ha)} for the less-luminous, 
mid O-type stars are taken at face value, they yield values that are 
inconsistent with the non-detection of wind profiles in the resonance 
lines of {\ion{P}{5}}, at least for the assumed solar abundance of P.
Although the magnitude of the discrepancy is indeterminate for these objects, 
the sense is the same as for the rest of the mid-O stars.  
The {\em largest} value of {\mdotq(\ilevel{P}{4+})} for {\em any} main 
sequence O star in our sample is {$7.2\times10^{-8}$ \msunpyr}, and most 
are less than this by an order of magnitude or more.
Thus, if $q$(\ilevel{P}{4+}) ever approaches unity for main 
sequence O stars, then their mass-loss rates are {\em much} smaller 
than theoretical expectations, which are $\gtrsim 10^{-7}$~{\msunpyr} 
for these stars \citep{Vink00}.

\subsection{Comparison of {\mdot(radio)} with {\mdot(\ha)}}\label{compare}

Although not a primary motivation of the present study, the compilation of
measurements of {\mdot} data in Table~\ref{mdots} provides an opportunity
to re-examine the agreement between measurements of {\mdot(radio)} and
{\mdot(\ha)}.
Detections are available for both diagnostics for a subset of 8 objects.  
A comparison of the two measurements for these stars indicates a systematic 
difference between them, with the mean value of
{ {\mdot(\ha)} / {\mdot(radio)} $\approx 2.4 \pm 0.7$}.   
This is contrary to previous studies \citep[e.g.,][]{Lamers93,Puls96}, 
which found agreement between the two measures.  
\citet{Blomme03} noted a similar -- though smaller -- discrepancy between 
{\mdot(radio)} and {\mdot(\ha)} determinations for HD~66811 [O4~I(n)f].

\section{Discussion}\label{discuss}

It has been known for some time {\citep[see, e.g.,][]{Hamann80}} that 
self-consistent wind models produce too much absorption in the {\ion{P}{5}} 
resonance lines of individual O-types stars.  
Although the precise range of spectral types where $q$(\ilevel{P}{4+})
is dominant is poorly defined at present, the large sample considered here
indicates that there is a significant, systematic discrepancy between 
mass-loss determinations for {\em all} objects with spectral types in
{\em either} of the two possible ranges discussed in \S\ref{p5range}.
Thus, the ``\ion{P}{5} problem" is not limited to a few, possibly peculiar
objects.
Instead, the discordance between {\mdot} diagnostics indicates that one or 
more biases must be influencing the analyses in such a way that either 
\mdotq(\ilevel{P}{4+})} is systematically {\em under-estimated}, or 
that {\mdot(radio)} and {\mdot(\ha)} are systematically {\em over-estimated}.  
A combination of biases might also be affecting these diagnostics by
different amounts.  
In the following sections, we examine the biases that might affect 
mass-loss rates determined by different methods within the context of the 
standard wind model (i.e., a smooth, steady, and spherically symmetric wind).  

\subsection{Biases in Measurements of {\mdot(\ha)}}

A measurement of {\mdot(\ha)} is complicated, because it requires
(a) accurate knowledge of the distribution of wind density in the
    near-star environment; 
(b) non-LTE calculations to determine the excitation equilibrium of H; and
(c) an accurate model for the underlying photospheric feature, which must
    also incorporate the blend with {\ion{He}{2}~$\lambda$6560}.
The latter two requirements are both sensitive to temperature,
with hotter temperatures usually leading to larger values of {\mdot}
\citep{Puls96}.  
Consequently, values of {\mdot(\ha)} may be systematically over-estimated
with old, hotter temperature scale.
However, since only one of the older \mdot(\ha) measurements was  
used in Figs.~\ref{cf_fig} and \ref{pv_teff} (see \S\ref{ha_mdot}), 
this is unlikely to be an important bias. 
If anything, the reduction in the temperature scale should move
values of {\mdot(\ha)} closer to {\mdotq(\ilevel{P}{4+})}.

A more subtle problem occurs for stars of luminosity class V--III, which 
typically exhibit a partially filled absorption profile rather than emission 
above the local continuum.  
For these stars, estimates of {\mdot(\ha)} depend very sensitively on 
how well the underlying photospheric profile is reproduced by the adopted 
model atmosphere.  
It is possible that small, systematic uncertainties in the strength of 
the photospheric {\ha} profiles contribute to the large 
discrepancy between {\mdotq(\ilevel{P}{4+})} and {\mdot(\ha)}
exhibited by the group of 5 mid-range O dwarfs and giants seen 
in Fig.~\ref{cf_fig}.

Another potential source of bias in determinations of {\mdot(\ha)} is the 
high frequency of variability \citep{Ebbets82,Kaper98,Markova05}.  
Many of these variations are characterized by changes in the equivalent width 
of the emission, which implies that either the amount of material or its 
density distribution changes.  
Consequently, it is difficult to evaluate the extent to which a measurement 
based on a single profile at a snapshot in time represents the 
global, time-averaged mass flux.  
Wind-wind interactions in short-period binary systems also produce 
complicated variations in {\ha}; see, e.g., the series of papers
bounded by \citet{Gies91} and \citet{Thaller01}.
Although the effects of colliding winds are not included in the analyses 
used to compile Table~\ref{mdots}, there is no particular evidence that that 
binaries depart from the general trends evident in Fig.~\ref{cf_fig}.

\subsection{Biases in Measurements of {\mdot(radio)}}

The {\mdot(radio)} are usually assumed to be the most reliable, since the 
physical mechanism responsible for the radio emission (free-free emission) 
is well understood; ionization corrections are generally insignificant 
{\citep{Lamers93}}; and contamination from the stellar photosphere is 
completely negligible.  
Furthermore, the radio photosphere is typically sufficiently far from the 
star that the details of the velocity law are not important: only {\vinf} 
is required, which can be determined accurately from UV resonance lines. 
In the context of the standard wind model, the only other input 
to {\mdot(radio)} is the distance.  
Although imperfect knowledge of the distances to Galactic O stars 
introduces scatter into the derived {\mdot(radio)}, the uncertainties 
should be random and are unlikely to result in a systematic bias.

However, it has been recognized for many years that the {\mdot(radio)} 
can be systematically over-estimated if a fraction of the observed radio 
flux is due to nonthermal emission.  
Approximately 25\% of O-type stars exhibit nonthermal radio emission, which 
can be recognized either through dramatic variability between epochs 
or from the frequency dependence of their radio spectra 
\citep{Bieging89}.  
Binarity may play a role in generating nonthermal electrons through wind-wind 
interactions; see, e.g., the recent discussion of radio observations of 
HD~93129A by \citet{Benaglia04}.  
However, other mechanisms may contribute to the nonthermal radio emission 
from the winds of single stars; see, e.g., \citet{VanLoo04}.  
Of the 16 stars in our sample with {\mdot(radio)} measurements, only 3 are 
considered ``definite" thermal emitters (HD~66811, HD~37742, and HD~152408) 
by \citet{Lamers93}, and 2 more are considered ``probable" thermal emitters 
(HD~30614 and HD~151804).  
The remaining 11 stars have measurements at a single frequency, which is 
insufficient to establish that the emission is thermal.  
Five of these objects are known or suspected binaries, although these stars 
do not appear to have systematically larger {\mdot(radio)} values compared 
to similar, single stars.  
Most of the stars with {\mdot(radio)} measurements also have 
{\mdot(\ha)} measurements that are larger still; see \S\ref{compare}.
Thus, it seems that any biases common to ``$\rho^2$" diagnostics affect 
{\mdot(\ha)} more than {\mdot(radio)}.

\subsection{Biases in Measurements of {\mdotq(\ilevel{P}{4+})}}

Equation~(\ref{eq_mdotq}) shows that \mdotq(\ilevel{P}{4+}) will be 
systematically under-estimated if either the oscillator strength, 
$f_{ij}$, or the P abundance, $A_P$, is over-estimated. 
Although random errors in the other parameters required to evaluate 
equation~(\ref{eq_mdotq}) will introduce uncertainties, the resulting
effects are unlikely to be systematic.

Since {\ilevel{P}{4+}} is a lithium-like ion, it is quite unlikely that its 
oscillator strength is uncertain by a substantial amount.  
\cite{Morton03} does not indicate that there is any controversy 
related to the value of $f_{ij}$.
Consequently, there is no reason to suspect that the oscillator 
strength biases measurements of {\mdotq(\ilevel{P}{4+})}.

The Galactic P abundance is more problematic.
Some authors {\citep[e.g.,][]{Pauldrach94,Pauldrach01}} have adopted 
subsolar values (e.g., P/P$_\sun$ of 0.50--0.67 and 0.05 for HD~66811 and 
HD~30614, respectively) in order to achieve good fits for wind profiles in 
the {\ion{P}{5}} resonance doublet.  
However, there is little evidence to support a systematically reduced 
abundance of P for Galactic O-type stars.
The solar P abundance by number relative to H is $A_P = 12.00 + 
\log( N_P / N_ H) = 5.45 \pm 0.06$ \citep{Biemont94}), which is $\sim$30\% 
smaller than the meteoric abundance {\citep[$A_P = 5.56$;~][]{Anders89}}.  
\citet{Dufton86} measured $A_P = 5.59$ in the Galactic interstellar medium 
from a survey of the {\ion{P}{2}}~$\lambda\lambda$1153, 1302 resonance doublet 
along 51 sight lines and found that it was only modestly depleted (factors of
3 or less) in cold clouds. 
More recently, \citet{Lebouteiller05} have also confirmed that the
interstellar abundance of P is solar.
Similarly, analysis of a {\fuse} spectrum of HD~207538 (B0~V, though a 
candidate for a chemically peculiar star) by \citet{Catanzaro03} shows that 
the photospheric lines of {\ion{P}{4}} are well reproduced with solar 
abundances. 
Thus, there is no particular evidence that the abundance of P is 
systematically subsolar in the material from which O stars form, or 
in their cooler, B-type relatives.
It is in principle possible to search for anomalous P abundances more 
explicitly by determining the photospheric abundance of {\ion{P}{5}} from 
O-type stars with weak winds (e.g., HD~47839; see Fig.~\ref{PV_O7fits}).

A more speculative explanation for the systematically small values of 
{\mdotq(\ilevel{P}{4+})} is that $q$(\ilevel{P}{4+}) peaks for  
O stars at a value that is substantially less than 1. 
Of course, this behavior is not predicted by standard wind models,
and would require that some unusual circumstances dominate the ionization
balance of P.
A candidate mechanism noted by \citet{Pauldrach94} is the approximate
coincidence of the ground-state ionization threshold of {\ion{P}{5}}
(65.023 eV; 191~\AA) with that of {\ion{He}{2}} (54.416 eV; 228~\AA).
As a result, \citet{Pauldrach94} speculated that $q$(\ilevel{P}{4+}) 
is driven by the behavior of {\ilevel{He}{+}}.  
To the best of our knowledge, this idea has not been developed further.
Another candidate mechanism is the production of soft X-rays, which
is already beyond the strict constraints of the ``standard model,"
but which could alter the ionization balance in the wind.
However, there is no evidence that ions with observable wind lines and 
ionization states that lie above and below {\ilevel{P}{4+}} -- e.g., 
{\ilevel{S}{5+}} and {\ilevel{P}{3+}} -- are strongly over-populated;
see the Appendix~\ref{App} for a plausibility argument in the case of HD~66811.

We conclude that, in the context of the standard wind model, there is no 
compelling reason why measurements of {\mdotq(\ilevel{P}{4+})} should be 
systematically under-estimated by factors of $\ge 10$.

\subsection{Relaxing the Standard Model \label{relax}}

The preceding discussion does not clearly identify any processes that 
could account for the systematic discrepancy of {\mdot} measurements 
made from what should otherwise be reliable diagnostics.  
Instead, it would seem that the discordance is an artifact of one or more 
of the assumptions inherent to the ``standard wind model."

The weakest assumption is that the wind has a smooth density distribution.  
The rampant variability of UV and {\ha} wind profiles 
\citep{Prinja86,Kaper96,Kaper98,deJong01,Markova05}; 
extended ``black troughs" and variable blue wings in saturated absorption 
troughs of P~Cygni profiles \citep{Lucy82,Puls93}; 
stochastically variable sub-structure in the emission-line profiles 
of HD~66811 \citep{Eversberg98}; 
and the detection of X-rays distributed through the wind 
\citep{Harnden79,Seward79,Chlebowski89,Cassinelli01}
all denote the presence of inhomogeneous structures on a variety of 
spatial scales.  
Theoretical calculations \citep{Owocki88,Feldmeier97} also indicate 
that line-driven stellar winds are subject to strong instabilities, 
which should redistribute wind material into dense structures.

\subsubsection{The effect of clumping on ``$\rho^2$"
               diagnostics\label{relax_rho2}}

The presence of clumping substantially alters the interpretation
of ``$\rho^2$" diagnostics like free-free continuum emission and {\ha} 
line emission.
Since the emission results from the interaction of two particles,
it will be produced more strongly from denser regions.
Consequently, if a clumped wind is interpreted in terms of 
the ``standard model" (which is smooth and homogeneous),
``$\rho^2$" diagnostics will necessarily {\em over-estimate} the 
true mass flux, since the excess emission produced in the inhomogeneities
will be incorrectly interpreted as arising from a smooth but denser medium.
Smooth but aspherical redistributions of wind material such as those
caused by rapid rotation \citep[see, e.g.,][]{Petrenz96} also
lead to over-estimates of the mass-loss rate by ``$\rho^2$" diagnostics,
again because the standard model provides an incorrect density distribution. 

Of course, the sensitivity of ``$\rho^2$" diagnostics to clumping has been 
recognized for a long time.  
It had been previously discounted on the basis of the perceived agreement
between {\mdot(\ha)} and {\mdot(radio)}, which was taken to indicate
that either the wind is clumped by the same amounts over huge distances
(which was viewed as unlikely) or it is not significantly 
clumped anywhere \citep{Lamers93}.
The reassessment of these mass-loss rates, which is partially driven 
by the lowering of the {\teff} scale for O-type stars, has challenged
the observational basis for this conclusion \citep{Repolust04}. 

Clumping is frequently described in terms a two-component model that consists 
of dense clumps characterized by density $\rho_H$ and a rarefied inter-clump 
gas with density $\rho_L$, which redistributes the material of a smooth 
flow while preserving its mean density, $\rho_s$:
\begin{equation}
   \rho_s = f \rho_H + (1 - f)\,\rho_L = \rho_H \left( f + 
    \left[ 1 - f \right] x  \right)
\end{equation}
where 
$f$ is the volume filling factor of the denser component, $ 0 < f \le 1$;
$x$ is the density contrast, $x = \rho_{L} / \rho_{H}$; and
where all quantities are understood to be functions of position.
Although initial applications of this formulation emphasized continuum
processes \citep[see, e.g.,][]{Abbott81}, it has also been used 
characterize the effect of clumping on line processes
\citep{Hillier98,Crowther02,Bouret05}.
Since the physical mechanism responsible for redistributing wind material
into dense clumps is not known, most implementations minimize the number
of free parameters by further assuming that the dense clumps are separated
by vacuum, so that $x=0$ and $\rho_s = f \rho_H$.
To put the present discussion on the same footing, we also follow this 
approach.

As shown by \citet{Abbott81}, the two-component model for clumping predicts 
that the {\mdot} determined from measurements of ``$\rho^2$" diagnostics, 
\mdot($\rho^2$), will be over-estimated by a factor of
\begin{equation}
 \left( {\left[ f + x\,( 1 - f )\right]^2} \over { f + (1 - f)x^2} 
 \right)^{1 \over 2}. 
\end{equation}
In the extreme case of dense clumps separate by vacuum, $ x = 0 $ and
{\mdot} is over-estimated in a smooth-wind model by $1/\sqrt{f}$; i.e., 
\mdot$(\rho^2)_{c} = \sqrt{f}\,$ \mdot$(\rho^2)_{s}$,
where the subscripts ``c" and ``s" denote ``clumped" and ``smooth," 
respectively.
In effect, this simplified model for clumping predicts a degeneracy between
$f$ and {\mdot$(\rho^2)$}; see \citet{Hillier99}.

\subsubsection{The effect of clumping on UV resonance lines \label{relax_uv}}

In contrast to ``$\rho^2$" diagnostics, determinations of {\mdotq} from the 
wind profiles of UV resonance lines are quite insensitive to clumping.
The analysis of these profiles is essentially a determination of the
optical depth (or, equivalently, the column density) of all the material
associated with a specific ion along the line of sight that causes
the observed P~Cygni absorption trough.
Since optical depth (or column density) is an {\em integral} quantity, these 
measurements are not sensitive to the distribution of material along the 
line of sight.  

Consequently, as long as the clumps remain optically thin on the spatial scales 
relevant to line transfer, no material will be hidden and measurements will 
reliably account for all of it irrespective of its distribution.
Thus, from equation~(\ref{eq_mdotq}), 
(\mdotq)$_s$ = (\mdotq)$_c$. 
The situation becomes more complicated if the clumps become optically thick,
since their shape and distribution must then be specified in order 
to determine the degree to which the wind is ``porous."
\citet{Owocki04} provide a useful formalism to describe porosity in this
context, and discuss how it can affect the predicted mass-loss rates.
\cite{Massa03} describe how porosity can affect the formation of 
P~Cygni lines and, in extreme cases, produce an apparently unsaturated
profile for a line that would be extremely saturated if the wind material
were distributed smoothly.

On larger spatial scales, inhomogeneities will be directly observable as 
significant departures from the expected shape of the P~Cygni absorption 
trough.
The discrete absorption components (DACs) that are nearly ubiquitous
in wind profiles of O-type stars are notable examples of such features; 
see, e.g., \citet{Prinja86,Howarth89,Kaper96,Kaper99}.
The SEI-fitting procedure described in \S\ref{uv_mdot} will characterize the
optical depth of large-scale structures like DACs reliably.

The detailed model-atmosphere analysis of HD~190429A (O4 If+) by
\citet{Bouret05} provides a test of the robustness of SEI fits to
small-scale clumping.
Our SEI fit to the {\ion{P}{5}} profiles in this star is excellent
(Fig.~\ref{PV_fits}), and consequently {\mdotq(\ilevel{P}{4+})} is 
very well determined. 
Similarly, \citet{Bouret05} achieved an excellent fit, but {\em only} 
for a ``clumped'' model with a P abundance reduced to 
P/P$_\sun = 0.5$; see their Fig.~2. 
Using this value of P/P$_\sun$, their {\mdot} (from their Table~4), 
and their mean {$q$(\ilevel{P}{4+}) = 0.5} over the velocity range 
{100--2000~\kms} (estimated from their Fig.~5), we find that the clumped
{\tt CMFGEN} model predicts {$\log$\mdotq(\ilevel{P}{4+}) = $-6.08$}.
For comparison, scaling the results of our SEI profile fitting by their
reduced P abundance gives {$\log$\mdotq(\ilevel{P}{4+}) = $-6.06$}.
This remarkable agreement confirms that both techniques determine the 
same optical depth in the line when clumping is incorporated in the models 
and all other factors are equal; i.e., clumping does not bias the 
determination of {\mdotq} from wind-profile fits to UV resonance lines.

However, even though the robustness of wind-profile fitting against clumping 
ensures that {\mdotq} will be determined reliably, we expect that $q$ will 
be affected by the presence of significant clumps in the wind.
Two-body interactions occur more frequently in denser clumps, so 
recombination to lower states is favored.
In general,
\begin{equation}
\dot M_c  = {q_s( {\rm P^{4+}} ) \over q_c( {\rm P^{4+}} )}\,\dot M_s~.
\end{equation}
The ratio $Q \equiv q_s( {\rm P^{4+}} ) / q_c( {\rm P^{4+}} )$ 
can be greater or less than 1, depending on whether $q_s$ is dominant 
($q_c$ decreases with respect to $q_s$; $Q > 1$ ) or one stage below dominant 
($q_c$ increases with respect to $q_s$; $Q < 1$).
For example, the models of HD~190429A by \citet[][(their Fig.~5]{Bouret05} 
show that in the high-velocity region of the wind, 
$q_s \approx 0.1$ and $q_c \approx 0.6$; i.e., $Q\approx 0.16$.
$Q$ can in principle be exactly 1 if clumping causes the gain in 
{$q$(\ilevel{P}{4+})} from {\ilevel{P}{5+}} to balance the loss to 
{\ilevel{P}{3+}}.
Unfortunately, it is not possible to determine the relevant populations
of these ions in a purely empirical fashion, though we show in 
Appendix~\ref{App} that some constraints can be inferred for the special 
case of HD~66811 ($\zeta$~Puppis}).

\subsection{Reconciling Mass-Loss Estimates}

The different behavior of ``$\rho^2$" diagnostics and the resonance lines
of dominant ions in the presence of small-scale clumping implies that
concordance between these measurements can in principle be achieved by
allowing for inhomogeneities in the wind.
In particular, measurements of {\mdot({\ion{P}{5})} can be used to break the 
degeneracy between {\mdot($\rho^2$)} and $f$ for stars where 
$q$(\ilevel{P}{4+}) $\sim$ 1.\footnote{\citet{Bouret05} noted that 
the radiatively pumped {\ion{O}{4}~$\lambda\lambda1338, 1343$} and 
{\ion{O}{5}~$\lambda 1371$} wind lines also help to break the 
{\mdot($\rho^2$)}--$f$ degeneracy.}
In terms of the simplified model for clumping discussed in \S\ref{relax_rho2}, 
concordance between these two estimates can be achieved when
\mdot(\ion{P}{5})$_c = $ \mdot$(\rho^2)_c\,$; i.e., when the filling factor is
\begin{equation}\label{eqn_f}
f = Q^2 \left( \dot{M}( {\rm P~V} )_s \over  \dot{M}(\rho^2)_s \right)^2~.
\end{equation}
Unfortunately, $Q$ is a function of both the degree of clumping and {\teff}.
Although this information is well defined in terms of the parameters of
specific model atmosphere computations, its behavior is difficult to predict
{\em a priori}.
Thus, for illustrative purposes, we arbitrarily assign $Q$ values between
0.1 and 10.
We expect this range to cover most cases where a {\ion{P}{5}} wind 
profile is observed, irrespective of whether smooth wind models predict 
{\ilevel{P}{4+}} to be dominant ($q_s \approx 1$; $Q > 1$ ) or a level below 
dominant ($q_s \ll 1$; $Q < 1$).

Table~\ref{ffactors} shows the filling factors implied by these assumptions
for the two possible ranges of {\teff} discussed in \S\ref{p5range}.
If smooth wind models correctly predict the {\teff} range where
$q_s$(\ilevel{P}{4+}) $\approx$ 1, then Q $\ga$ 1; but if the hotter
range indicated by $q_{est}$ is appropriate, then $Q \la 1$.
In either case, Table~\ref{ffactors} shows that {\em substantial}
degrees of clumping are required to achieve concordance, irrespective 
of the adopted value of $Q$.

The filling factors associated with the simplified model of clumping
introduced in \S\ref{relax_rho2} have also been determined from comprehensive
spectroscopic analysis of individual objects.
To date, these analyses have been undertaken with {\tt CMFGEN} 
\citep{Hillier98}, which implements velocity-dependent filling factors that 
decrease exponentially from 1 (i.e., smooth) near the photosphere to a 
``terminal" filling factor, $f_\infty$, at great distances from the star.
Since the exponential decrease in $f$ is very rapid, the values of 
$f_\infty$ from these analyses should correspond to the mean 
values of $f$ implied by the mass-loss discrepancy.
The values of $f_\infty$ derived from detailed spectroscopic fits to O stars 
in the Small Magellanic Cloud are typically $\sim$0.1 
\citep{Crowther02,Hillier03,Bouret03,Evans04}.
\citet{Bouret05} found that clumped models with values of  $f_\infty$ = 0.04 
and 0.02 were required to fit the UV spectra of two Galactic 
O stars, HD~190429A (O4~If$+$) and HD~96715 [O4~V((f))], respectively.
Although these values of $f$ are not as extreme as the illustrative values
provided in Table~\ref{ffactors}, we conclude that qualitatively, both 
detailed spectroscopic modeling and a straightforward analysis of the 
mass-loss discrepancy indicate that the winds of at least some O stars are 
strongly clumped.

Of course, it remains to be seen whether such small filling factors are 
physically realistic.
The application of the simplified model of clumping described in 
\S\ref{relax_rho2}} to line transfer has not been justified rigorously.
Although it likely captures some of the effects associated with density 
enhancements, it is fundamentally inappropriate for line processes because 
it does not specify a length scale.
As a result, the clumps are assumed to remain optically thin irrespective 
of their density enhancement, which is unlikely to be appropriate
for very small filling factors (which imply very large values
of $\rho_H/\rho_s$). 
In addition, the usual approximation for the density contrast (i.e.,
$x=0$) is aphysical (``Nature abhors a vacuum"\footnote{Attributed 
to Fran\c{c}ois Rabelais (c. 1494--1553), a French monk, satirist, and 
physician.}, etc.), and leads to extreme values of $f$.
Finally, observations of stellar-wind profiles indicate that large-scale
structures are generally present, and these are likely to contribute their
own perturbations to the assumed density structure in addition to 
small-scale clumping.
For all these reasons, the filling factors derived from the simplest version
of the ``two-component" model are likely to be too small.

\section{Conclusions}\label{conclude}

Comparison of \mdotq(\ilevel{P}{4+}) values determined from wind-profile
fitting with mass-loss rates determined from radio and {\ha} emission 
reveals a large, systematic discrepancy between diagnostics that should
be reliable mass-loss estimators. 
The magnitude of this discrepancy depends crucially on the range of {\teff} 
for which {\ilevel{P}{4+}} is the dominant ion in the wind. 
If, on the one hand, the predictions of the ionization balance of P based 
on smooth-wind models are adopted, then {\ilevel{P}{4+}} is dominant 
between O7.5 and O9.7. 
In this case, the median discrepancy is enormous:
\mdot($\rho^2$)/\mdot(\ion{P}{5}) $\approx$ 130. 
If, on the other hand, the ``$\rho^2$" diagnostics are taken at face 
value, then $q_{\rm est}$(\ilevel{P}{4+}) is observed to peak for
temperature classes between O4 and O7 at values of $\sim$11\% for 
supergiants, 6\% for giants, and {$\sim$4\% for dwarfs.
The fact that these maximal values are not unity underscores the 
discordant estimates of {\mdot}.
The median discrepancy over this subset of our sample is more modest:
\mdot($\rho^2$)/\mdot(\ion{P}{5}) $\approx$ 20.
\citet{Massa03} found similar behavior for $q_{\rm est}$(\ilevel{P}{4+}) in 
a sample of O stars in the Large Magellanic Cloud.

Thus, irrespective of the range of {\teff} for which 
$q$({\ilevel{P}{4+}}) $\approx$ 1, we conclude that there is a significant 
discrepancy in {\mdot} estimates for some Galactic O stars.
We interpret the disagreement between these diagnostics as a
signature of the presence of significant, small-scale clumping in 
the winds of some O-type stars (and, by likely extension, all O stars, 
though perhaps to substantially different degrees).
If the winds are clumped, then the values of {\mdot} measured from ``$\rho^2$" 
diagnostics are over-estimated, while the values of {\mdotq} measured
from a dominant ion like {\ilevel{P}{4+}} will be modestly affected, if they
are affected at all.
On the basis of a simple model for clumping, we find that consistent
estimates of the mass-loss rate can be recovered if the dense material
is characterized by volume filling factors that are very small.

Since previous determinations and calibrations of {\mdot} have relied
on ``$\rho^2$" diagnostics, we conclude that mass-loss rates have 
been over-estimated for at least some O-type stars.
Thus, the primary consequence of the present study is that estimates
of the mass-loss rate must be decreased substantially, typically by factors
of $\sim$100 [if $q$({\ilevel{P}{4+}}) $\sim$ 1 between O7.5 and O9.7] 
or $\sim$20 [if $q$({\ilevel{P}{4+}}) $\sim$ 1 between O4 and O7].
Even larger decreases are implied for mid-range dwarfs.
Interestingly, independent studies of the X-ray emission from 
O stars -- in particular HD~36486A 
\citep[$\delta$~Orionis A; O9.5~II;~][]{Miller02} 
and $\zeta$~Pup \citep{Kramer03} -- have found that the 
cool wind contributes significantly less attenuation than expected for the 
adopted mass-loss rates, which also suggests that {\mdot} has been 
over-estimated.
Revisions to {\mdot} of the magnitude suggested by the present study will 
have wide-ranging consequences for both the evolution of massive stars and 
the feedback they provide to their interstellar environments.

Our conclusions concerning the importance of clumping are similar to those 
reached in studies of the winds of Magellanic O stars 
\citep{Crowther02,Massa03,Hillier03,Bouret03,Evans04};
a survey of luminous Galactic B stars \citep{Prinja05}; 
and the detailed analysis of two Galactic O4 stars \citep{Bouret05}.
All these studies highlight the utility of unsaturated resonance lines of 
dominant or near-dominant ions as mass-loss indicators.
The \ion{P}{5} resonance lines are the crucial diagnostics for O-type stars;
\citet{Evans04} have suggested that the \ion{S}{4}~$\lambda\lambda$1073, 1074
doublet plays an analogous role for spectroscopic analysis of late-O and
early B-type supergiants.
Access to these lines must now be considered essential for detailed 
spectroscopic studies of early-type stars.

At an even more fundamental level, this work indicates that the 
``standard model" for hot-star winds needs to be amended, since predictions
based on it appear to be inadequate for many purposes.
In particular, clumping needs to be incorporated realistically as a  
general feature.
In order to make this feasible, the mechanisms responsible for the formation 
of clumps must to be understood, as must the evolution of these 
inhomogeneities under the ambient radiative and hydrodynamic forces.
The implication is that knowledge of the time-dependent behavior of 
line-driven stellar winds on large- and small-spatial scales is required 
to understand even the basic properties of these outflows, and in particular 
to interpret their observational signatures correctly.

\acknowledgements
We wish to thank the referee, Rolf Kudritzki, for his detailed comments 
on this paper. 
Many improvements resulted from his insightful criticisms.
We are likewise grateful to Joachim Puls for fruitful discussion of many of 
the issues explored here.
This work is based in part on observations made with the NASA-CNES-CSA Far 
Ultraviolet Spectroscopic Explorer, which is operated by the Johns Hopkins 
University for NASA under contract NAS5-32985.  
Financial support from NASA Research Grants grants associated with FUSE 
Guest Investigator program E082 (NNG04EH17P to SGT, Inc. and NNG04GM49G to 
JHU) is also gratefully acknowledged.

{\it Facilities:} \facility{FUSE}, \facility{Copernicus}, \facility{Orfeus(BEFS)}.

\appendix
\section{Empirical Constraints on the Ionization of P in the Wind of
         HD~66811}\label{App}

Ideally, the ionization state of P in a stellar wind could be probed 
directly through the ratios
$q$(\ilevel{P}{5+})/$q$(\ilevel{P}{4+}) and 
$q$(\ilevel{P}{3+})/$q$(\ilevel{P}{4+}).
Unfortunately, the resonance lines of {\ilevel{P}{5+}} lie near
{90~\AA} and are inaccessible. 
The resonance line of {\ilevel{P}{3+}} falls at {951~\AA}, which is 
accessible to {\fuse} but almost always compromised by extinction or 
blending; see, e.g., the far-UV atlases of \citet{Pellerin02}
and \citet{Walborn02}.

Instead, {\ilevel{S}{3+}} and {\ilevel{S}{5+}} can be used as surrogates
for {\ilevel{P}{3+}} and {\ilevel{P}{5+}} to constrain the ionization 
state of P. 
Although these matches are not perfect, they are useful indicators
of the prevailing ionization conditions in the winds of mid-O stars, 
because
(a) the ionization ranges of the surrogates correspond approximately, 
    and in any case bracket, the range of {\ilevel{P}{4+}} 
    \citep[see, e.g.,][their Fig.~1]{Massa03};
    and
(b) like P, S is a non-CNO element whose abundance does not change
    over the life of a given star, and is unlikely to exhibit gross
    abundance differences between stars.
Owing to its modest reddening, HD~66811 [$\zeta$~Puppis; O4~I(n)f]
is one of the few Galactic O stars for which the {\ion{S}{6}} resonance
doublet can be measured.
Consequently, the following plausibility argument is based on 
the {\copernicus} spectrum of $\zeta$~Pup \citep{Morton77}.
The salient characteristics of this spectrum are that the {\ion{S}{4}} wind 
lines are weaker than those of {\ion{P}{5}}, while the wind lines 
of {\ion{S}{6}} are strong but not saturated.

To connect the ionization properties of P and S, consider that 
for two wind lines of the same star (where the $w(x)$ and {\mdot}
are identical for all lines), equation~(\ref{eq_mdotq}) gives
\begin{equation}
\frac{q^{(1)}}{q^{(2)}} = \frac{\tau_{rad}^{(1)}}{\tau_{rad}^{(2)}} \times
       \frac{\lambda_{0}^{(2)} f_{ij}^{(2)} A_{E}^{(2)}}
            {\lambda_{0}^{(1)} f_{ij}^{(1)} A_{E}^{(1)}} \;\;\;.
\end{equation}
Since the products of $\lambda_0 f_{ij} A_E$\/ are known for each resonance 
line and the values of $\tau_{rad}$ can be determined by wind-profile 
fitting, the relative strengths of wind lines can be used to infer the 
relative ion fractions.
For solar abundances, the ratios of {$\lambda_0 f_{ij} A_E$\/} for 
{\ion{S}{4}/\ion{P}{5}} and {\ion{S}{6}}/{\ion{P}{5}} are approximately
6 and 59, respectively.
Then, by using the observed S ions as surrogates for the unobserved P ions:
\begin{enumerate}

\item The observation that the {\ion{S}{4}} doublet is weaker than the 
      {\ion{P}{5}} doublet in the spectrum of HD~66811 implies that 
      $\tau_{rad}$({\ion{S}{4}})/$\tau_{rad}$({\ion{P}{5}}) $\le$ 1, and hence
      {$q$(\ilevel{P}{4+}) $>$ 6\,$q$(\ilevel{S}{3+})} $\approx 
      6\,q$(\ilevel{P}{3+}).

\item The fact that the {\ion{S}{6}} doublet is not saturated implies that
      $\tau_{rad}$(\ion{S}{6}) $\lesssim 5$, while
      $\tau_{rad}$(\ion{P}{5}) $\simeq 1$.
      Hence $q$(\ilevel{P}{4+}) $\gtrsim$ 12\,$q$(\ilevel{S}{5+})
      $\approx 12\,q$(\ilevel{P}{5+}).
      
\end{enumerate}
Thus, in the wind of HD~66811, we infer that $q$(\ilevel{P}{4+})
is greater than the ion fractions of either of its adjacent states.
While this does not in itself prove that $q$(\ilevel{P}{4+}) $\approx$1,
it shows that it is not possible to hide significant amounts of P
as {\ilevel{P}{3+}} or {\ilevel{P}{5+}}, even if the wind is clumped.


%
%
\onecolumn
\newpage
\begin{figure}
   \plotone{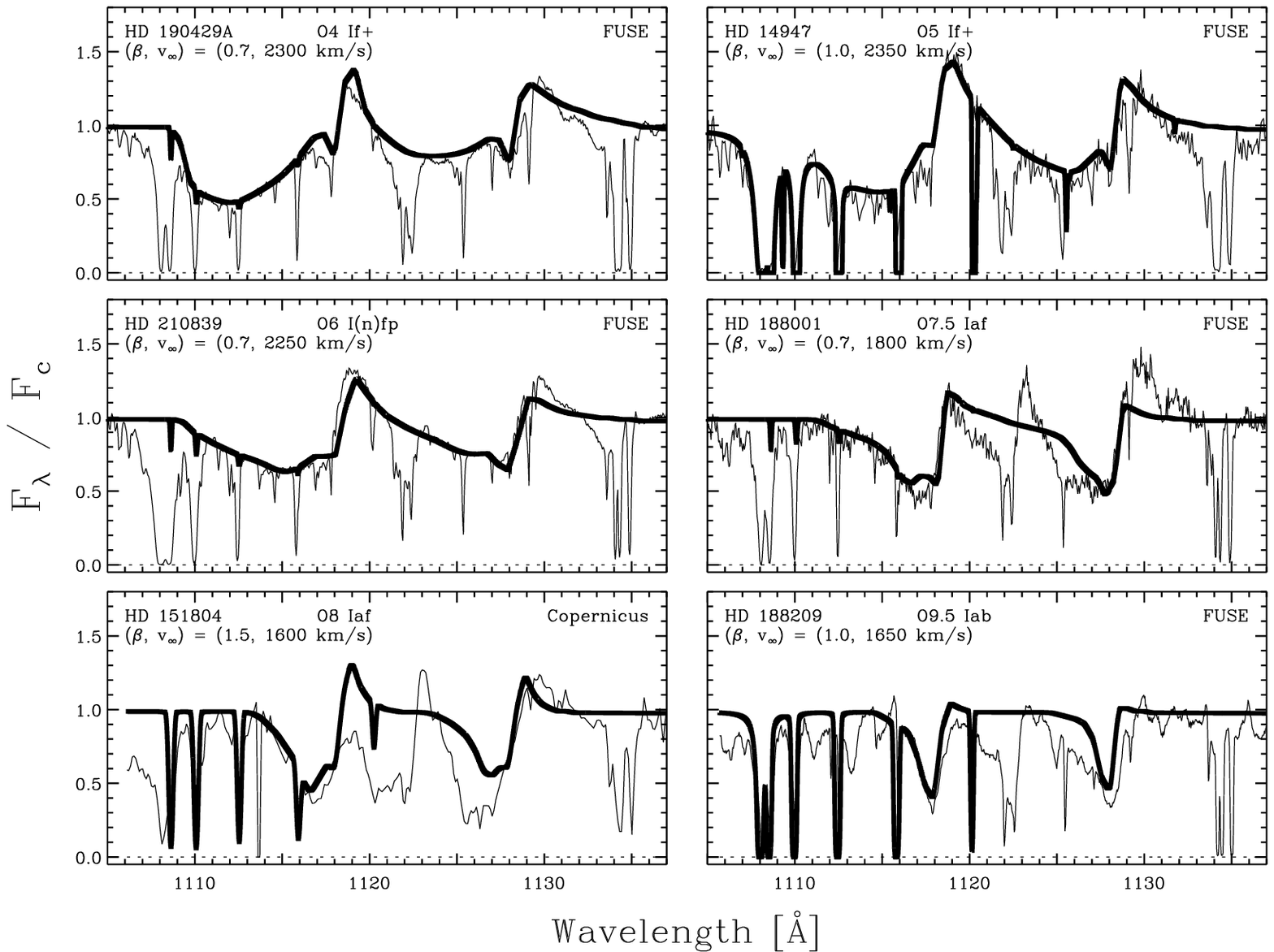}
   \figcaption{
      Examples of fits to the {\ion{P}{5}} wind profiles of O4--O9.5
      supergiants.
      Since the red component of the doublet is increasingly contaminated
      by the blend with {\ion{Si}{4}}~$\lambda$1128, the fit to the  blue
      component is given greatest weight for spectral types later than 
      $\sim$O7. \label{PV_fits}
     }
\end{figure}
\newpage
\begin{figure}
   \plotone{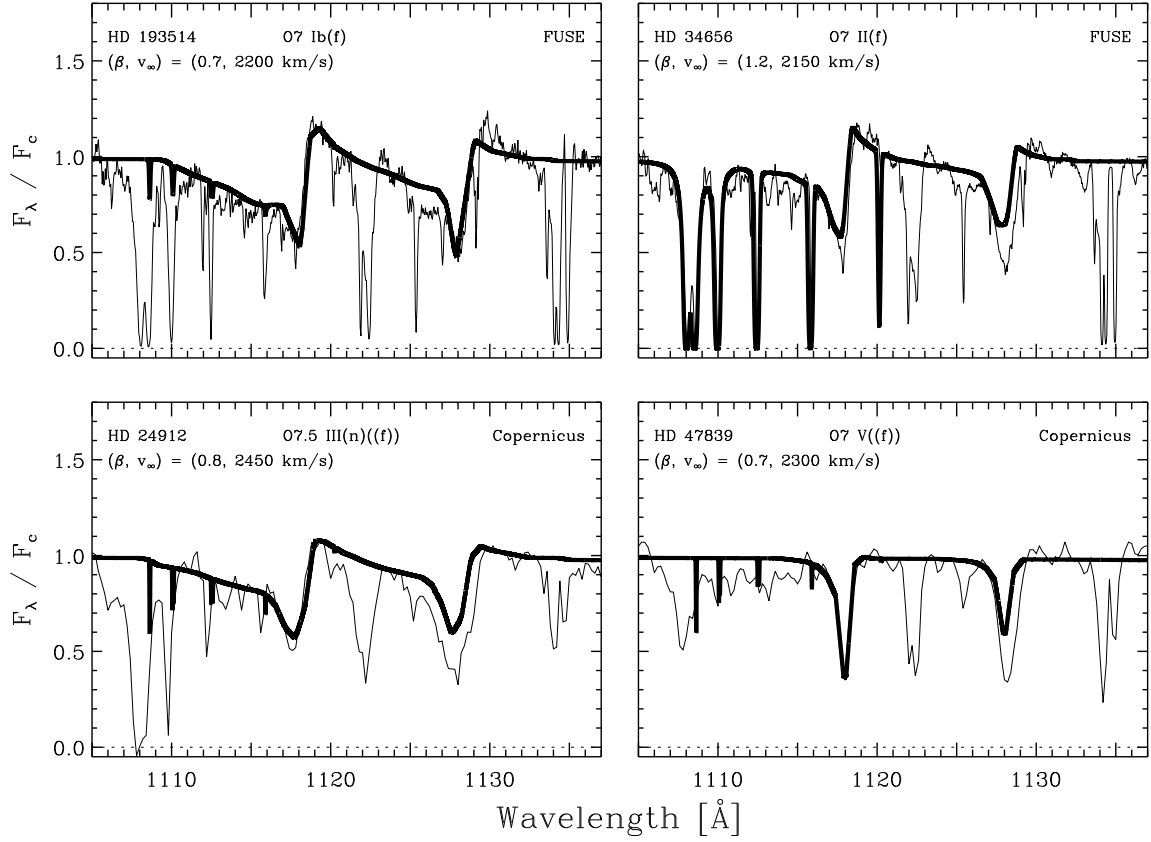}
   \figcaption{
      Examples of fits to the {\ion{P}{5}} wind profiles of luminosity
      class I--V stars with temperature class of O7.  
      In all cases, the red component of the doublet is blended with
      {\ion{Si}{4}}~$\lambda$1128. 
      The fits to HD~47839 only provide upper limits to the wind contribution
      to the {\ion{P}{5}} wind profiles, which are dominated by the underlying
      photospheric component. \label{PV_O7fits}
      }
\end{figure}
\newpage
\begin{figure}               
   \plotone{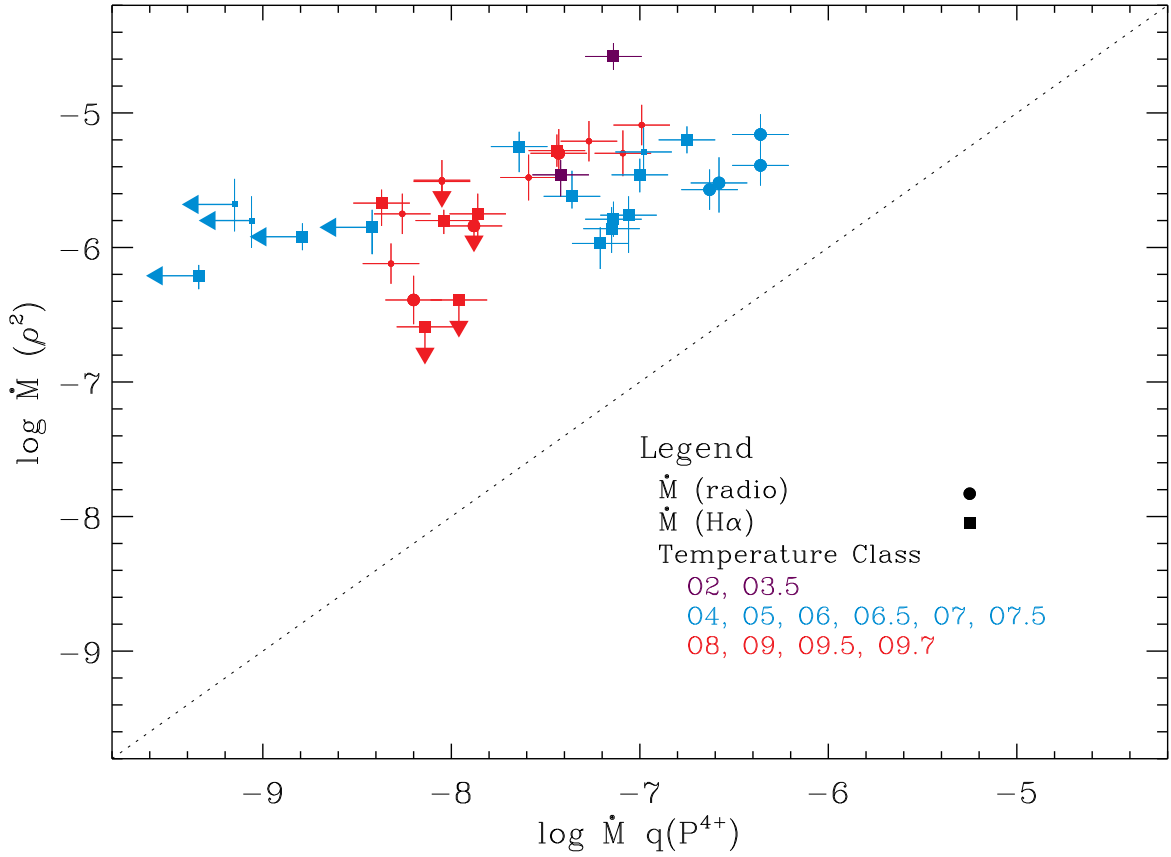}
   \figcaption{
    Comparison of {\mdot} with {\mdotq(\ilevel{P}{4+})}. 
    The shapes of symbols distinguish radio (circles) and
    {\ha} measurements, while symbol size separates the primary (large) 
    and secondary (small) samples. 
    Upper limits on non-detections are indicated by arrows.
    Color coding divides the entire sample into early-
    (O2, O3, O3.5), mid- (O4--O7.5), and late-O types (O8--O9.7).
    The dotted line denotes a 1--1 correlation between the two 
    measurements. \label{cf_fig}
    }
\end{figure}
\newpage
\begin{figure}
   \plotone{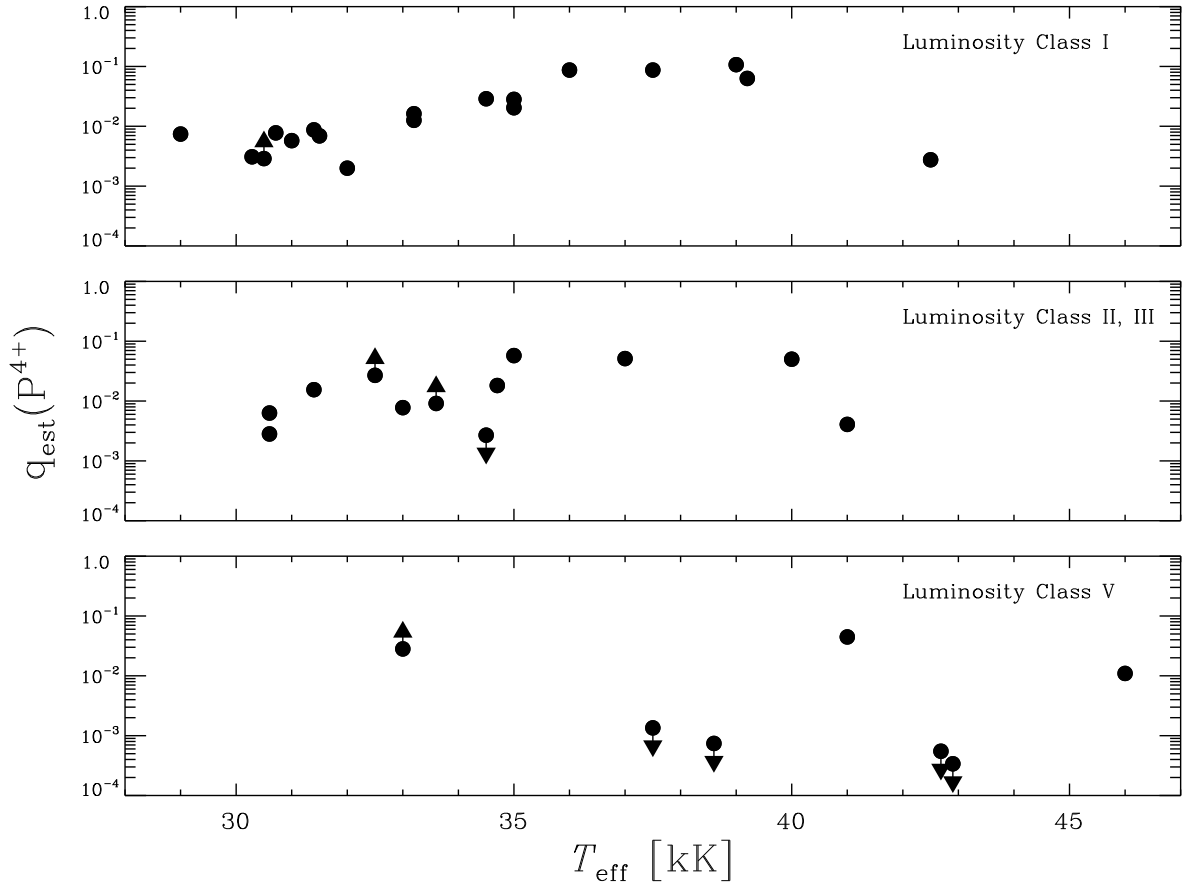}
   \figcaption{
    $q_{\rm est} \equiv \dot M\,q({\rm P^{4+}}) / \dot M( \rho^2 )$
    as a function of {\teff} for different luminosity classes.
    Non-detections of {\ion{P}{5}} are indicated by downward-pointing arrows, 
    while upward-pointing arrows represent {\ion{P}{5}} detections divided by 
    upper limits on non-detected radio or {\ha} emission. 
    The upper limit at {\teff}$=$33~kK in panel for dwarfs is for HD~13268,
    which is a very rapid rotator with a nitrogen-rich atmosphere. 
    \label{pv_teff}
    }
\end{figure}
%
\begin{deluxetable}{llcccccccccccl}
\tabletypesize{\scriptsize}
\tablecaption{Galactic O-Star Sample\label{targets}}
\tablecolumns{14}
\tablewidth{0pt}
\tablehead{ \colhead{HD/HDE}                                  &
            \multicolumn{2}{c}{Spectral Type}                 &
	    \colhead{~~~}                                     &
	    \multicolumn{3}{c}{Derived Parameters}            &
	    \colhead{~~~}                                     &
	    \colhead{\vsini~\tablenotemark{a}}                &
	    \colhead{~~~}                                     &	    
	    \multicolumn{2}{c}{\vinf}                         &
	    \colhead{~~~}                                     &
	    \multicolumn{1}{l}{Notes\tablenotemark{b}}        \\
	    \cline{2-3} \cline{5-7} \cline{9-9} \cline{11-12} \\
	    \colhead{}                                        & 
	    \colhead{Classification}                          & 	   
	    \colhead{Ref.}                                    & 
	    \colhead{}                                        & 	    
 	    \colhead{\teff~[kK]}                              &  
 	    \colhead{\rsun}                                   & 
	    \colhead{Ref.}                                    & 	    
	    \colhead{}                                        & 	    
            \colhead{\kms}                                    &
	    \colhead{}                                        &
	    \colhead{\kms}                                    & 
    	    \colhead{Ref.}                                    & 
	    \colhead{}                                        &	    
	    \colhead{}                                         }	    
\startdata
\sidehead{Primary Sample}
\phn13268  & ON8 V            & \phn1 && 33.0 &    10.3 & \phn2 &&    309\phantom{:} &&    2150 & \phn3 &&                         \\
\phn14947  & O5 If+           & \phn4 && 37.5 &    16.8 & \phn2 &&    133\phantom{:} &&    2350 & \phn3 &&                         \\
\phn15558  & O5 III(f)        & \phn4 && 41.0 &    18.2 & \phn2 &&    123\phantom{:} &&    2800 & \phn3 &&  SB1O                   \\
\phn24912  & O7.5 III(n)((f)) & \phn4 && 35.0 &    14.0 & \phn2 &&    213\phantom{:} &&    2450 & \phn3 &&  $\xi$~Per              \\ 
\phn30614  & O9.5 Ia          & \phn4 && 29.0 &    32.5 & \phn2 &&    129\phantom{:} &&    1550 & \phn3 && $\alpha$~Cam            \\ 
\phn34656  & O7 II(f)         & \phn5 && 34.7 &    24.1 & \phn6 && \phn91\phantom{:} &&    2150 & \phn7 &&                         \\
\phn36861A & O8 III((f))      & \phn4 && 33.6 &    15.1 & \phn6 && \phn74\phantom{:} &&    2400 & \phn3 &&  $\lambda$~Ori A        \\
\phn37043A & O9 III           & \phn4 && 31.4 &    21.6 & \phn6 &&    116\phantom{:} &&    2450 & \phn8 &&  $\iota$~Ori; SB2O      \\
\phn42088  & O6.5 V           & \phn5 && 38.6 &    10.7 & \phn6 && \phn65\phantom{:} &&    2200 & \phn3 &&                         \\	
\phn47839  & O7 V((f))        & \phn4 && 37.5 & \phn9.9 & \phn6 && \phn67\phantom{:} &&    2300 & \phn8 &&  SB1O                   \\
\phn66811  & O4 I(n)f         & \phn5 && 39.0 &    19.4 & \phn2 &&    219\phantom{:} &&    2250 & \phn9 &&  $\zeta$~Pup            \\
\phn93129A & O2 If*           &    10 && 42.5 &    22.5 & \phn2 &&    180\phantom{:} &&    3200 & \phn3 &&  SB                     \\ 
\phn93250  & O3.5 V((f+))     &    10 && 46.0 &    15.9 & \phn2 &&    107:           &&    3250 & \phn3 &&                         \\   
   149757  & O9.5 Vnn         &    11 && 32.0 & \phn8.9 & \phn2 &&    372\phantom{:} &&    1500 &    12 &&  $\zeta$~Oph            \\
   188209  & O9.5 Iab         & \phn5 && 31.0 &    19.6 & \phn6 && \phn92\phantom{:} &&    1650 & \phn7 &&                         \\
   190429A & O4 If+           & \phn4 && 39.2 &    20.8 & \phn6 &&    105\phantom{:} &&    2300 & \phn8 &&                         \\
   190864  & O6.5 III(f)      & \phn4 && 37.0 &    12.3 & \phn2 && \phn88\phantom{:} &&    2500 & \phn3 &&  SB1?                   \\      
   191423  & O9 III:n         & \phn4 && 32.5 &    12.9 & \phn2 &&    436\phantom{:} &&    1150 & \phn9 &&                         \\
   192639  & O7 Ib(f)         & \phn5 && 35.0 &    18.7 & \phn2 && \phn96\phantom{:} &&    2150 & \phn3 &&                         \\
   193514  & O7 Ib(f)         & \phn5 && 34.5 &    19.3 & \phn2 && \phn94\phantom{:} &&    2200 & \phn3 &&                         \\
   193682  & O4: III(f)       &    13 && 40.0 &    13.1 & \phn2 && \nodata           &&    2800 & \phn9 &&                         \\
   203064  & O7.5 III:n((f))  & \phn5 && 34.5 &    15.7 & \phn2 &&    305\phantom{:} &&    2550 & \phn9 &&  68~Cyg                 \\
   207198  & O9 Ib-II         & \phn5 && 33.0 &    16.6 & \phn2 && \phn91\phantom{:} &&    2150 & \phn3 &&                         \\
   209975  & O9.5 Ib          & \phn5 && 32.0 &    22.9 & \phn2 && \phn95\phantom{:} &&    2050 & \phn3 &&  19~Cep                 \\  
   210809  & O9 Iab           & \phn4 && 31.5 &    21.2 & \phn2 &&    117\phantom{:} &&    2100 & \phn3 &&                         \\
   210839  & O6 I(n)fp        & \phn4 && 36.0 &    21.1 & \phn2 &&    219\phantom{:} &&    2250 & \phn3 &&  $\lambda$~Cep          \\ 
   217086  & O7 Vn            & \phn4 && 36.0 & \phn8.6 & \phn2 &&    332\phantom{:} &&    2550 & \phn3 &&                         \\  
   303308  & O4 V((f+))       &    10 && 41.0 &    11.5 & \phn2 &&    111\phantom{:} &&    3100 & \phn3  &&  SB1?                  \\ 
                                                                                                                                   \\  
\sidehead{Secondary Sample}													  
\phn36486A & O9.5 II          & \phn5 && 30.6 &    17.7 &    14 &&    144\phantom{:} &&    2000 &    15 &&  $\delta$~Ori A; SB1OE  \\ 
\phn37742  & O9.7 Ib          & \phn5 && 30.5 &    22.1 &    14 &&    124\phantom{:} &&    2100 & \phn8 &&  $\zeta$~Ori            \\ 
\phn46150  & O5 V((f))        &    10 && 40.9 &    11.2 &    14 &&    111\phantom{:} &&    2900 & \phn8 &&  SB2?                   \\ 
\phn46223  & O4 V((f+))       &    10 && 42.9 &    12.4 &    14 && \phn82\phantom{:} &&    2800 & \phn8 &&                         \\
\phn57061  & O9 II            & \phn5 && 30.6 &    17.7 &    14 &&    120\phantom{:} &&    1960 &    15 &&  $\tau$~CMa; SBE        \\
   149038  & O9.7 Iab         & \phn5 && 30.5 &    22.1 &    14 && \phn86\phantom{:} &&    1750 & \phn8 &&  $\mu$~Nor              \\
   149404  & O9 Ia            & \phn4 && 31.4 &    21.8 &    14 &&    100\phantom{:} &&    2450 &    15 &&  SB2O                   \\
   151804  & O8 Iaf           & \phn4 && 33.2 &    21.1 &    14 &&    104\phantom{:} &&    1600 & \phn8 &&                         \\
   152408  & O8: Iafpe        & \phn5 && 33.2 &    21.1 &    14 && \phn85\phantom{:} && \phn960 &    15 &&                         \\   
   152424  & OC9.7 Ia         & \phn5 && 30.5 &    22.1 &    14 && \phn86\phantom{:} &&    1760 &    15 &&  SB1?                   \\  
   164794  & O4 V((f))        & \phn4 && 42.9 &    12.4 &    14 && \phn70\phantom{:} &&    2950 & \phn8 &&  9~Sgr; SB2?            \\
   188001  & O7.5 Iaf         & \phn5 && 35.0 &    20.5 &    14 && \phn93\phantom{:} &&    1800 & \phn8 &&  9~Sge; SB1?            \\    
\enddata                                                                                                       
\tablenotetext{a}{From \citet{Howarth97}.}
\tablenotetext{b}{Binary status: SB1 -- single-line spectroscopic binary; SB2 -- double-line spectroscopic binary.
                  Suffixes: O -- orbit derived; E -- eclipsing system; ? -- possible binary.}
\tablerefs{ (1)~\citealt{Mathys89};
            (2)~\citealt{Repolust04};
	    (3)~\citealt{Haser95};
	    (4)~\citealt{Walborn73};
	    (5)~\citealt{Walborn72};
	    (6)~\citealt{Markova04};
	    (7)~\citealt{Howarth97};
	    (8)~\citealt{Groenewegen89a};
	    (9)~\citealt{Puls96};
	   (10)~\citealt{Walborn02};
	   (11)~\citealt{Lesh68};
	   (12)~this work;
	   (13)~\citealt{Garmany91};
	   (14)~\citealt{Martins05};
	   (15)~\citealt{Prinja90}. }
\end{deluxetable}
\newpage
\begin{deluxetable}{llcllcrcccl}
\tabletypesize{\scriptsize}
\tablecaption{Log of Far-UV Observations\label{obslog}}
\tablecolumns{11}
\tablewidth{0pt}
\tablehead{ \colhead{HD/HDE}                       &
            \colhead{Spectral Type}                &
	    \colhead{Obs.~\tablenotemark{a}}       &
	    \colhead{Data Set~\tablenotemark{b}}   &
	    \colhead{PI}                           &
	    \colhead{Date Obs.~\tablenotemark{c}}  &
	    \colhead{Int. Time~\tablenotemark{d}}  &
	    \colhead{N(exp.)~\tablenotemark{e}}    &
	    \colhead{Mode~\tablenotemark{f}}       &
	    \colhead{Aperture}                     &
	    \colhead{Notes}                         }
\startdata	    
\sidehead{Primary Sample}
\phn13268  & ON8 V            & F & P1020304 & Jenkins    & 1999-11-24 &   4438 & \phn2 & TTAG & LWRS    &                   \\
\phn14947  & O5 If+           & F & E0820201 & Massa      & 2004-09-30 &   5614 & \phn3 & TTAG & LWRS    &                   \\  
\phn15558  & O5 III(f)        & F & P1170101 & Hutchings  & 1999-11-28 &   2472 & \phn4 & HIST & LWRS    &                   \\
\phn24912  & O7.5 III(n)((f)) & C & C017-001 & \nodata    & 1972-10-30 &  36.32 &    18 & U2   & \nodata &                   \\
\phn30614  & O9.5 Ia          & C & C021-001 & \nodata    & 1972-11-09 &  51.59 &    29 & U2   & \nodata &                   \\
\phn34656  & O7 II(f)         & F & P1011301 & Savage     & 2000-03-01 &   4179 & \phn7 & HIST & LWRS    &                   \\
\phn36861A & O8 III((f))      & C & C026-003 & \nodata    & 1972-11-27 &  79.96 &    32 & U2   & \nodata &                   \\
\phn37043A & O9 III           & C & C033-005 & \nodata    & 1972-12-25 &  70.01 &    28 & U2   & \nodata &                   \\ 
\phn42088  & O6.5 V           & F & P1021101 & Jenkins    & 2000-11-05 &   4219 & \phn9 & HIST & LWRS    &                   \\
\phn47839  & O7 V((f))        & C & C028-004 & \nodata    & 1972-12-04 &  28.58 &    19 & U2   & \nodata &                   \\
\phn66811  & O4 I(n)f         & C & C044-001 & \nodata    & 1973-02-22 &  71.91 &    37 & U2   & \nodata &                   \\
\phn93129A & O2 If*           & F & P1170202 & Hutchings  & 2000-01-27 &   7371 & \phn2 & TTAG & LWRS    &                   \\
\phn93250  & O3.5 V((f+))     & F & P1023801 & Jenkins    & 2000-02-04 &   4140 & \phn4 & HIST & LWRS    &                   \\
   149757  & O9.5 Vnn         & C & C002-004 & \nodata    & 1972-09-05 &  48.51 &    42 & U2   & \nodata &                   \\   
   188209  & O9.5 Iab         & F & S5231106 & Sahnow     & 2004-10-08 &  11542 &    24 & HIST & HIRS    & \tablenotemark{g} \\
   190429A & O4 If+           & F & P1028401 & Jenkins    & 2000-07-18 &   5390 &    10 & HIST & LWRS    &                   \\
   190864  & O6.5 III(f)      & F & E0820501 & Massa      & 2004-05-23 &   3468 & \phn3 & TTAG & LWRS    &                   \\
   191423  & O9 III:n         & F & E0821301 & Massa      & 2004-06-09 &   4795 & \phn4 & TTAG & LWRS    &                   \\
   192639  & O7 Ib(f)         & F & C1710101 & Nichols    & 2002-09-04 &  14720 & \phn7 & TTAG & LWRS    &                   \\
   193514  & O7 Ib(f)         & F & E0820701 & Massa      & 2004-06-08 &   6478 & \phn2 & TTAG & LWRS    &                   \\
   193682  & O4: III(f)       & F & E0820301 & Massa      & 2004-06-09 &   5818 & \phn5 & TTAG & LWRS    &                   \\
   203064  & O7.5 III:n((f))  & C & C152-004 & \nodata    & 1979-07-08 &  23.29 &    13 & U2   & \nodata &                   \\
   207198  & O9 Ib-II         & F & P1162801 & Snow       & 2000-07-23 &  13180 & \phn3 & TTAG & LWRS    &                   \\
   209975  & O9.5 Ib          & F & D0140302 & Federman   & 2004-07-26 &   1294 & \phn3 & HIST & LWRS    & \tablenotemark{h} \\ 
   210809  & O9 Iab           & F & P1223103 & Jenkins    & 2000-08-08 &  10065 &    18 & HIST & LWRS    &                   \\
   210839  & O6 I(n)fp        & F & P1163101 & Snow       & 2000-07-22 &   6050 &    10 & HIST & LWRS    &                   \\
   217086  & O7 Vn            & F & E0820801 & Massa      & 2004-07-25 &   4970 & \phn2 & TTAG & LWRS    &                   \\
   303308  & O4 V((f+))       & F & P1221602 & Jenkins    & 2000-05-27 &   7692 &    12 & HIST & LWRS    &                   \\
                                                                                                                          \\
\sidehead{Secondary Sample}
\phn36486A & O9.5 II          & C & C025-001 & \nodata    & 1972-11-21 &  70.76 &    48 & U2   & \nodata &                   \\
\phn37742  & O9.7 Ib          & C & C024-004 & \nodata    & 1972-11-18 &  71.94 &    44 & U2   & \nodata &                   \\
\phn46150  & O5 V((f))        & F & P1021401 & Jenkins    & 2001-03-05 &   4888 & \phn9 & HIST & LWRS    &                   \\
\phn46223  & O4 V((f+))       & F & C1680302 & Bruhweiler & 2004-02-23 &   7122 &    14 & HIST & LWRS    &                   \\
\phn57061  & O9 II            & C & C046-001 & \nodata    & 1973-03-16 &  71.37 &    43 & U2   & \nodata &                   \\ 
   149038  & O9.7 Iab         & F & P3032601 & Williger   & 2004-09-11 &     58 & \phn1 & HIST & LWRS    & \tablenotemark{h}  \\
   149404  & O9 Ia            & F & P1161702 & Snow       & 2001-08-07 &  17850 &    38 & HIST & LWRS    &                   \\
   151804  & O8 Iaf           & C & C065-002 & \nodata    & 1974-09-05 &  13.47 &    11 & U2   & \nodata &                   \\
   152408  & O8: Iafpe        & C & C067-002 & \nodata    & 1974-09-04 &   9.43 & \phn8 & U2   & \nodata &                   \\
   152424  & OC9.7 Ia         & F & Z9016301 & Andersson  & 2002-04-18 &   7313 & \phn2 & TTAG & LWRS    &                   \\
   164794  & O4 V((f))        & B & BEFS1037 & \nodata    & 1993-09-16 &    257 & \phn1 & TTAG & \nodata &                    \\
   188001  & O7.5 Iaf         & F & E0820901 & Massa      & 2004-05-29 &     58 & \phn1 & HIST & LWRS    & \tablenotemark{h} \\
\enddata
\tablenotetext{a}{Observatory used to obtain spectra: B = {\it BEFS}; C={\copernicus}; F={\fuse}.}
\tablenotetext{b}{The identifier for the data set in MAST. {\copernicus} data sets are coadded scans.}
\tablenotetext{c}{UT date when the scan or integration started, expressed as year-month-day.}
\tablenotetext{d}{For {\copernicus} spectra: the time betwen the start of the first scan and the end of the last scan, in hours.
                  For {\fuse} and {BEFS} spectra: the total integration time, in s.}
\tablenotetext{e}{For {\copernicus} spectra: number of scans in coadded file. 
                  For {\fuse} amd {BEFS} spectra: number of exposures comprising the total integration.}
\tablenotetext{f}{For {\copernicus} spectra: the detector. For {\fuse} and {BEFS} spectra: the mode of detector operation.}
\tablenotetext{g}{FP-SPLIT observation; part of a test for observing bright objects.}
\tablenotetext{h}{{LiF1} only observation.}		   
\end{deluxetable}
\newpage
\begin{deluxetable}{llccccccccr}
\tabletypesize{\scriptsize}
\tablecaption{Rescaled Measurements of {\mdot(radio)}\label{revmdot}}
\tablecolumns{11}
\tablewidth{0pt}
\tablehead{ \colhead{HD}                            &
            \colhead{Spectral Type}                 &
	    \multicolumn{3}{c}{Previous}            &
	    \colhead{~~}                            & 
            \multicolumn{5}{c}{Revised}             \\
	    \cline{3-5} \cline{7-11}                \\
	    \colhead{}                              &
            \colhead{}                              & 
            \colhead{$d_{old}$}                     & 
	    \colhead{$\log$ \mdot}                  & 
            \colhead{Ref.}                          & 
	    \colhead{}                              & 
            \colhead{$M_V$}                         & 
	    \colhead{Ref.}                          & 
	    \colhead{$d_{new}$}                     & 
	    \colhead{$f_{corr}$}                    & 
	    \colhead{$\log$ \mdot}                  \\ 
	    \colhead{}                              &
            \colhead{}                              & 
            \colhead{(kpc)}                         & 
	    \colhead{(\msunpyr)}                    & 
            \colhead{}                              & 
	    \colhead{}                              & 
            \colhead{}                              & 
	    \colhead{}                              & 
	    \colhead{(kpc)}                         & 
	    \colhead{}                              & 
	    \colhead{(\msunpyr)}                       } 
\startdata
\sidehead{Primary Sample}
\phn14947   & O5 If+      &  2.0 & $-5.52$ & 1 && $-5.94$ & 2 & 2.0 & 1.0 & $    -5.52$ \\
\phn30614   & O9.5 Ia     &  1.0 & $-5.41$ & 3 && $-7.00$ & 2 & 1.2 & 1.3 & $    -5.30$ \\
\phn36861A  & O8 III((f)) &  0.5 & $-6.04$ & 3 && $-5.85$ & 4 & 0.7 & 1.6 & $\le -5.84$ \\
\phn37043A  & O9 III      &  0.5 & $-6.50$ & 3 && $-6.24$ & 4 & 0.6 & 1.3 & $    -6.39$ \\
\phn66811   & O4 I(n)f    &  0.4 & $-5.54$ & 3 && $-6.32$ & 2 & 0.5 & 1.4 & $    -5.39$ \\
149757      & O9.5 Vnn    &  0.1 & $-7.41$ & 3 && $-4.35$ & 2 & 0.2 & 2.8 & $    -6.96$ \\
190429A     & O4 If+      &  1.7 & $-5.34$ & 1 && $-6.51$ & 4 & 2.2 & 1.5 & $    -5.16$ \\
210839      & O6 I(n)fp   &  0.8 & $-5.65$ & 3 && $-6.40$ & 2 & 0.9 & 1.2 & $    -5.57$ \\ \\
\sidehead{Secondary Sample}
\phn36486A  & O9.5 II     &  0.5 & $-5.97$ & 3 && $-5.73$ & 5 & 0.4 & 0.7 & $    -6.12$ \\
\phn37742   & O9.7 Ib     &  0.5 & $-5.60$ & 3 && $-6.28$ & 5 & 0.4 & 0.7 & $    -5.75$ \\
\phn57061   & O9 II       &  1.5 & $-5.20$ & 3 && $-5.77$ & 5 & 0.9 & 0.5 & $    -5.50$ \\
149038      & O9.7 Iab    &  1.3 & $-5.36$ & 3 && $-6.28$ & 5 & 1.0 & 0.7 & $\le -5.51$ \\
149404      & O9 Ia       &  1.4 & $-4.91$ & 3 && $-6.29$ & 5 & 0.9 & 0.5 & $    -5.21$ \\
151804      & O8 Iaf      &  1.9 & $-5.00$ & 3 && $-6.30$ & 5 & 1.2 & 0.5 & $    -5.30$ \\
152408      & O8: Iafpe   &  1.9 & $-4.87$ & 3 && $-6.30$ & 5 & 1.4 & 0.6 & $    -5.09$ \\
152424      & O7.5 Iaf    &  1.9 & $-5.26$ & 3 && $-6.28$ & 5 & 1.4 & 0.6 & $    -5.48$ \\
\enddata
\tablerefs{ (1)~\citealt{Scuderi98};
            (2)~\citealt{Repolust04};
            (3)~\citealt{Lamers93};
            (4)~\citealt{Markova04};
            (5)~\citealt{Martins05}. }
\end{deluxetable}

\newpage
\begin{deluxetable}{llccl}
\tabletypesize{\scriptsize}
\tablecaption{Properties of the \ion{P}{5} Resonance Doublet \label{pfive}}
\tablecolumns{5}
\tablewidth{0pt}
\tablehead{ \multicolumn{2}{l}{Property}              &
            \colhead{Value}                           &
	    \colhead{Ref.}                            &
	    \colhead{Remarks}                          } 
\startdata
Rest Wavelength                       & $ \lambda_{0,blue}, \lambda_{0,red}$ &  1117.977~\AA, 1128.008~\AA  & 1   &                   \\
Oscillator Strength                   & $ f_{blue}, f_{red}                $ &  0.473, 0.234                & 1   &                   \\
Abundance by number with respect to H & $ 12.00 + \log( N_P / N_ H)        $ &  5.45 $\pm$ 0.06             & 2   & Solar Photosphere \\
\enddata
\tablerefs{(1)~\citealt{Morton03};
           (2)~\citealt{Biemont94}. } 
\end{deluxetable}
\newpage
\begin{deluxetable}{llcrccrcccrccc}
\tabletypesize{\scriptsize}
\tablecaption{Derived and Adopted Mass-Loss Rates\label{mdots}}
\tablecolumns{13}
\tablewidth{0pt}
\setlength{\tabcolsep}{0.02in}
\tablehead{ \colhead{HD/HDE}                                         &
            \colhead{Spectral Type}                                  &
	    \colhead{$\beta$}                                        &
	    \multicolumn{2}{c}{$\log${\mdotq(\ilevel{P}{4+})}~\tna } &
	    \colhead{}                                               &
	    \multicolumn{3}{c}{$\log${\mdot}(radio)~\tna }           &
	    \colhead{}                                               &
	    \multicolumn{3}{c}{$\log${\mdot}(\ha)~\tna}              \\
            \cline{4-5} \cline{7-9} \cline{11-13}                    \\
	    \colhead{}                                               &
	    \colhead{}                                               &
	    \colhead{}                                               &
	    \colhead{$\log$\mdotq}                                   &
	    \colhead{$\sigma$}                                       &
	    \colhead{}                                               &
	    \colhead{$\log$\mdot}                                    &
	    \colhead{$\sigma$}                                       &
	    \colhead{Ref.}                                           &
	    \colhead{}                                               &
	    \colhead{$\log$\mdot}                                    &
	    \colhead{$\sigma$}                                       &
	    \colhead{Ref.}                                           &	    
	    }
\startdata	    
\sidehead{Primary Sample}
\phn13268  & ON8 V           & 0.7 & $    -8.14$ & $(-0.30, +0.18)$ &~~~&   \nodata   & \nodata              & \nodata &~~~& $\le -6.59$ & $(-0.34, +0.16)$ & 1       \\
\phn14947  & O5 If+          & 1.0 & $    -6.58$ & $(-0.12, +0.10)$ &   & $    -5.52$ & $(-0.22, +0.19)$\tnb & 2       &   & $    -5.07$ & $(-0.10, +0.10)$ & 1       \\  
\phn15558  & O5 III(f)       & 1.0 & $    -7.64$ & $(-0.30, +0.18)$ &   &   \nodata   & \nodata              & \nodata &   & $    -5.25$ & $(-0.19, +0.11)$ & 1       \\   
\phn24912  & O7.5 III(n)((f))& 0.8 & $    -7.21$ & $(-0.12, +0.10)$ &   &   \nodata   & \nodata              & \nodata &   & $    -5.97$ & $(-0.19, +0.12)$ & 1       \\   
\phn30614  & O9.5 Ia         & 1.0 & $    -7.43$ & $(-0.12, +0.10)$ &   & $    -5.30$ & $\pm 0.18$           & 3       &   & $    -5.22$ & $(-0.10, +0.10)$ & 1       \\  
\phn34656  & O7 II(f)        & 1.2 & $    -7.36$ & $(-0.12, +0.10)$ &   &   \nodata   & \nodata              & \nodata &   & $    -5.62$ & $(-0.09, +0.19)$ & 4       \\  
\phn36861A & O8 III((f))     & 0.7 & $    -7.88$ & $(-0.30, +0.18)$ &   & $\le -5.84$ & $\pm 0.18$           & 3       &   & $    -6.01$ & $(-0.10, +0.24)$ & 4       \\  
\phn37043A & O9 III          & 0.7 & $    -8.20$ & $(-0.30, +0.18)$ &   & $    -6.39$ & $\pm 0.18$           & 3       &   & $    -5.86$ & $(-0.10, +0.10)$ & 4       \\  
\phn42088  & O6.5 V          & 1.0 & $\le -9.34$ & \nodata          &   &   \nodata   & \nodata              & \nodata &   & $    -6.21$ & $(-0.10, +0.08)$ & 4       \\
\phn47839  & O7 V((f))       & 0.7 & $\le -8.79$ & \nodata          &   &   \nodata   & \nodata              & \nodata &   & $    -5.92$ & $(-0.10, +0.10)$ & 4       \\   
\phn66811  & O4 I(n)f        & 0.5 & $    -6.36$ & $(-0.12, +0.10)$ &   & $    -5.39$ & $\pm 0.15$           & 3       &   & $    -5.06$ & $(-0.13, +0.13)$ & 1       \\   
\phn93129A & O2 If*          & 1.0 & $    -7.14$ & $(-0.12, +0.10)$ &   &   \nodata   & \nodata              & \nodata &   & $    -4.58$ & $(-0.10, +0.10)$ & 1       \\   
\phn93250  & O3.5 V((f+))    & 1.0 & $    -7.42$ & $(-0.12, +0.10)$ &   &   \nodata   & \nodata              & \nodata &   & $    -5.46$ & $(-0.16, +0.11)$ & 1       \\  
   149757  & O9.5 Vnn        & 1.0 & $\le -9.88$ & \nodata          &   & $    -6.96$ & $\pm 0.16$           & 3       &   & $\le -6.75$ & $(-0.66, +0.17)$ & 1       \\   
   188209  & O9.5 Iab        & 1.0 & $    -8.04$ & $(-0.30, +0.18)$ &   &   \nodata   & \nodata              & \nodata &   & $    -5.80$ & $(-0.10, +0.08)$ & 4       \\     
   190429A & O4 If+          & 0.7 & $    -6.36$ & $(-0.12, +0.10)$ &   & $    -5.16$ & $(-0.16, +0.15)$\tnb & 2       &   & $    -4.85$ & $(-0.10, +0.08)$ & 4       \\  
   190864  & O6.5 III(f)     & 0.7 & $    -7.15$ & $(-0.12, +0.10)$ &   &   \nodata   & \nodata              & \nodata &   & $    -5.86$ & $(-0.18, +0.16)$ & 1       \\   
   191423  & O9 III:n        & 0.7 & $    -7.96$ & $(-0.30, +0.18)$ &   &   \nodata   & \nodata              & \nodata &   & $\le -6.39$ & $(-0.19, +0.11)$ & 1       \\   
   192639  & O7 Ib(f)        & 0.7 & $    -6.75$ & $(-0.12, +0.10)$ &   &   \nodata   & \nodata              & \nodata &   & $    -5.20$ & $(-0.10, +0.10)$ & 1       \\   
   193514  & O7 Ib(f)        & 0.7 & $    -7.00$ & $(-0.12, +0.10)$ &   &   \nodata   & \nodata              & \nodata &   & $    -5.46$ & $(-0.13, +0.12)$ & 1       \\  
   193682  & O4: III(f)      & 0.7 & $    -7.06$ & $(-0.12, +0.10)$ &   &   \nodata   & \nodata              & \nodata &   & $    -5.76$ & $(-0.28, +0.14)$ & 1       \\   
   203064  & O7.5 III:n((f)) & 1.0 & $\le -8.42$ & \nodata          &   &   \nodata   & \nodata              & \nodata &   & $    -5.85$ & $(-0.20, +0.13)$ & 1       \\ 
   207198  & O9 Ib-II        & 0.7 & $    -7.86$ & $(-0.30, +0.18)$ &   &   \nodata   & \nodata              & \nodata &   & $    -5.75$ & $(-0.17, +0.15)$ & 1       \\  
   209975  & O9.5 Ib         & 1.0 & $    -8.37$ & $(-0.30, +0.18)$ &   &   \nodata   & \nodata              & \nodata &   & $    -5.67$ & $(-0.17, +0.10)$ & 1       \\ 
   210809  & O9 Iab          & 0.7 & $    -7.44$ & $(-0.12, +0.10)$ &   &   \nodata   & \nodata              & \nodata &   & $    -5.28$ & $(-0.12, +0.12)$ & 1       \\  
   210839  & O6 I(n)fp       & 0.7 & $    -6.63$ & $(-0.12, +0.10)$ &   & $    -5.57$ & $\pm 0.15$           & 3       &   & $    -5.16$ & $(-0.16, +0.16)$ & 1       \\  
   217086  & O7 Vn           & 1.0 & $\le -8.67$ & \nodata          &   &   \nodata   & \nodata              & \nodata &   & $\le -6.64$ & $(-0.68, +0.19)$ & 1       \\   
   303308  & O4 V((f+))      & 0.7 & $    -7.14$ & $(-0.12, +0.10)$ &   &   \nodata   & \nodata              & \nodata &   & $    -5.79$ & $(-0.13, +0.13)$ & 1       \\  
                                                                                                                                                               \\
\sidehead{Secondary Sample}                                                                                                                                  
\phn36486A & O9.5 II         & 1.0 & $    -8.32$ & $(-0.30, +0.18)$ &   & $    -6.12$ & $\pm 0.15$           & 3       &   & \nodata     & \nodata          & \nodata \\ 
\phn37742  & O9.7 Ib         & 1.0 & $    -8.26$ & $(-0.30, +0.18)$ &   & $    -5.75$ & $\pm 0.15$           & 3       &   & \nodata     & \nodata          & \nodata \\
\phn46150  & O5 V((f))       & 1.0 & $\le -9.21$ & \nodata          &   &   \nodata   & \nodata              & \nodata &   & $\le -8.00$ & \nodata          & 5       \\
\phn46223  & O4 V((f+))      & 1.0 & $\le -9.06$ & \nodata          &   &   \nodata   & \nodata              & \nodata &   & $    -5.80$ & \nodata          & 5       \\  
\phn57061  & O9 II           & 1.0 & $    -8.05$ & $(-0.30, +0.18)$ &   & $    -5.50$ & $\pm 0.15$           & 3       &   & \nodata     & \nodata          & \nodata \\  
   149038  & O9.7 Iab        & 1.0 & $    -8.05$ & $(-0.30, +0.18)$ &   & $\le -5.51$ & $\pm 0.18 $          & 3       &   & \nodata     & \nodata          & \nodata \\
   149404  & O9 Ia           & 0.7 & $    -7.27$ & $(-0.12, +0.10)$ &   & $    -5.21$ & $\pm 0.15$           & 3       &   & \nodata     & \nodata          & \nodata \\  
   151804  & O8 Iaf          & 1.5 & $    -7.09$ & $(-0.12, +0.10)$ &   & $    -5.30$ & $\pm 0.17$           & 3       &   & \nodata     & \nodata          & \nodata \\    
   152408  & O8: Iafpe       & 1.5 & $    -6.99$ & $(-0.12, +0.10)$ &   & $    -5.09$ & $\pm 0.15$           & 3       &   & \nodata     & \nodata          & \nodata \\   
   152424  & OC9.7 Ia        & 0.7 & $    -7.59$ & $(-0.30, +0.18)$ &   & $    -5.48$ & $\pm 0.17$           & 3       &   & \nodata     & \nodata          & \nodata \\ 
   164794  & O4 V((f))       & 1.0 & $\le -9.15$ & \nodata          &   &   \nodata   & \nodata              & \nodata &   & $    -5.68$ & \nodata          & 5       \\
   188001  & O7.5 Iaf        & 0.7 & $    -6.98$ & $(-0.12, +0.10)$ &   &   \nodata   & \nodata              & \nodata &   & $    -5.29$ & \nodata          & 5       \\     
\enddata
\tablenotetext{a}{In units of \msunpyr.}
\tablenotetext{b}{The uncertainties quoted by \citealt{Scuderi98} have been increased by $\pm$0.10~dex to include uncertainties in the distance.}
\tablerefs{ (1)~\citealt{Repolust04}; 
            (2)~\citealt{Scuderi98}~\citetext{revised in Table~\ref{revmdot}}; 
	    (3)~\citealt{Lamers93}~\citetext{revised in Table~\ref{revmdot}}; 
	    (4)~\citealt{Markova04}; 
            (5)~\citealp{Puls96}~\citetext{their Table~11}. }
\end{deluxetable}
\newpage
\begin{deluxetable}{lccccc}
\tabletypesize{\scriptsize}
\tablecaption{Estimated Volume Filling Factors\label{ffactors}}
\tablecolumns{6}
\tablewidth{0pt}
\tablehead{ \colhead{Dominant Ion Predictor}                        &
            \colhead{Sp. Type Range}                                &	    
	    \colhead{{\mdot} Ratio \tablenotemark{a}}               &
            \multicolumn{3}{c}{Implied $f$}                         \\
	    \cline{4-6}                                             \\
	    \colhead{}                                              &
	    \colhead{}                                              &
	    \colhead{}                                              &
	    \colhead{$Q=0.1$}                                       &
	    \colhead{$Q=1  $}                                       &
	    \colhead{$Q=10 $}                                        }
\startdata
{\tt FASTWIND}                 & O7.5 -- O9.7 & 0.008 &      \nodata         & $6.4 \times 10^{-5}$  &  $6.4 \times 10^{-3}$ \\
 $q_{\rm est}$(\ilevel{P}{4+}) & O4   -- O7.5 & 0.050 & $2.5 \times 10^{-5}$ & $2.5 \times 10^{-3}$  & \nodata               \\
\enddata
\tablenotetext{a}{Median value of {\mdot(\ion{P}{5})$_s$}/{\mdot$(\rho^2)_s$} for the range of 
                  spectral types.}
\end{deluxetable}

%
\end{document}